\documentclass[a4paper,11pt]{article}
\usepackage[margin=2.5cm]{geometry}%to change margin width

\usepackage{setspace}
\usepackage{siunitx}
\usepackage[sort]{natbib}

\usepackage{hyperref}
\hypersetup{colorlinks = true, linktocpage = true, bookmarksopen = true, linkcolor = blue, urlcolor=blue, citecolor = blue, allcolors=blue}

\usepackage{tabu}
\usepackage{float}
\usepackage[ruled,vlined]{algorithm2e}
\usepackage{amssymb,amsmath,amsthm,enumitem}

\usepackage{svg} %include svg files
\usepackage{makecell} %used to make tables pretty

\renewcommand{\baselinestretch}{1.5}

\usepackage{verbatim} 
\usepackage{graphicx}
\usepackage{enumitem}

\usepackage[utf8]{inputenc}
\usepackage{amsmath}
\usepackage{amssymb}
\usepackage{bm}
\usepackage{tikz}
\usetikzlibrary{plotmarks}
\usepackage{float}
\usepackage{graphicx}

\newcommand{\edit}{\color{black}}
\usepackage[normalem]{ulem} %for strikeout text

\begin{document}

\renewcommand{\baselinestretch}{1.2}
   \begin{center}
      \Large\textbf{Strategic model reduction by analysing model sloppiness: a case study in coral calcification}\\
      \large{Sarah A. Vollert$^{a,b,c,*}$, Christopher Drovandi$^{a,b,c}$, Gloria M. Monsalve-Bravo$^{d,e,f}$ \& Matthew P. Adams$^{a,b,c,d}$} \\

      \footnotesize
$^a$Centre for Data Science, Queensland University of Technology, Brisbane, QLD 4001, Australia \\

$^b$ARC Centre of Excellence for Mathematical and Statistical Frontiers, Queensland University of Technology, Brisbane, QLD 4001, Australia \\

$^c$School of Mathematical Sciences, Queensland University of Technology, Brisbane, QLD 4001, Australia\\

$^d$School of Chemical Engineering, The University of Queensland, St Lucia, QLD 4072, Australia\\

$^e$School of Earth and Environmental Sciences, The University of Queensland, St Lucia, QLD 4072, Australia\\

$^f$Centre for Biodiversity and Conservation Science, The University of Queensland, St Lucia, QLD 4072, Australia\\

$^*$Corresponding author. E-mail: sarah.vollert@connect.qut.edu.au

   \end{center}

\section*{Highlights}
\begin{itemize}
\normalsize
\setlength{\parskip}{0cm}
    \item We propose a new model reduction framework based on a sloppy analysis.
    \item Model reduction is guided by identifying mechanisms which weakly inform predictions.
    \item We successfully reduce a complex coral growth model, preserving predictive accuracy.
    \item A Bayesian sloppy analysis is advantageous when the likelihood surface is complex.
\end{itemize}

\renewcommand{\baselinestretch}{1}
\setlength{\parskip}{1em}

\section*{Abstract}
\normalsize
It can be difficult to identify ways to reduce the complexity of large models whilst maintaining predictive power, particularly where there are hidden parameter interdependencies. Here, we demonstrate that the analysis of model sloppiness can be a new invaluable tool for strategically simplifying complex models. Such an analysis identifies parameter combinations which strongly and/or weakly inform model behaviours, yet the approach has not previously been used to inform model reduction. Using a case study on a coral calcification model calibrated to experimental data, we show how the analysis of model sloppiness can strategically inform model simplifications which maintain predictive power. Additionally, when comparing various approaches to analysing sloppiness, we find that Bayesian methods can be advantageous when unambiguous identification of the best-fit model parameters is a challenge for standard optimisation procedures. \\ \normalsize{Key words: \textit{sensitivity analysis, Bayesian inference, Sequential Monte Carlo, maximum likelihood estimation, parameter interdependence, model reduction}}

\newpage
\normalsize

\section{Introduction}
Mathematical models are used to understand complex biological and ecological systems, and to make predictions about system behaviours even in extreme conditions \citep{Getz_2018,Jeong_2018}. These models ideally include as much of the expected system dynamics as possible in order to gain mechanistic exploratory power \citep{Snowden_2017}.  However, this constructionist approach can lead to large, complex models, which can become problematic for various reasons. {\edit Complexity can introduce practical issues -- such as difficulty implementing, calibrating, solving, and interpreting models \citep{Hong_2017} -- as well as resulting in over-fitting, poor predictive performance, uncertainty in estimated parameter values and potentially becoming unscientific by being harder to disprove. Models should aim to balance being simple enough to capture general trends without over-fitting, but complex enough to capture the key features of a dataset -- this is known as the bias-variance tradeoff \citep{Geman_1992_biasvariance}.} 

Development of strategies to reduce a model’s complexity without loss of explanatory power is a research area of increasing interest \citep{Transtrum_2014_MBAM,Jeong_2018,hjelkrem_2017,Snowden_2017}. However, it can be challenging to identify appropriate model reductions for complex biological or ecological models, because there are many possible ways to reduce a model \citep{Jeong_2018,Cox_2006}. More so, parameter interdependencies can make it difficult to determine the informativeness of individual model components on the outputs \citep{Transtrum_2014_MBAM,Gibbons_2010}. Hence, systematic model reduction methods can be helpful for maintaining predictive power when simplifying a model. 

There are a variety of existing methods in the literature which propose ways of simplifying an underlying model. For example, projection-based methods aim to reduce the degrees of freedom of a model \citep{Schilders_2008_MOR}. {\edit Systematically removing parameters from models is challenging \citep{Transtrum_2014_MBAM}, so} mechanistic-focused methods commonly fix parameter values or state variables \citep{Cox_2006,Crout_2009,Elevitch_2020,hjelkrem_2017,Lawrie_2007} or lump them together \citep{Snowden_2017,pepiot2019model,Liao1988_lumping,Huang_2005_lumping} to make models more efficient and easier to calibrate. {\edit Sensitivity analysis methods are typically used to identify the importance of variations in parameter values on model outputs \citep{Saltelli_2004,Mara_2017}, and can inform model reductions in a factor-fixing setting} \citep{hjelkrem_2017,Van_2009,Cox_2006,Hsieh_2018}. {\edit Sensitivity analyses used for model reduction typically focus on the sensitivity of model outputs to changes in parameter values. However, we may instead be interested in which parameters are constrained by an available dataset. Additionally, the effects of changing model parameters on outputs often depend greatly on the assumed values of other parameters; hence we need to consider the effect of parameters in combination on model behaviours, rather than individuals \citep{gutenkunst_2007, brown_2003, Brown_2004, Gloria_2021_sloppy, Transtrum_2011,Transtrum_2014_MBAM}.}

This paper presents a new approach to strategic model reduction based on an analysis of model sloppiness. {\edit This type of sensitivity analysis captures the sensitivity of model outputs informed by a dataset; it looks at model-data fit sensitivities, rather than just model sensitivities.} {\edit The analysis of model sloppiness draws on dimension reduction techniques to} identify hidden parameter interdependencies (\citeauthor{Transtrum_2011}, \citeyear{Transtrum_2011}, \citeyear{Transtrum_2015}). Such parameter interdependencies, or parameter combinations, are identified by analysing the curvature of the surface which describes how the model-data fit depends on the model parameters \citep{Gloria_2021_sloppy,brown_2003,Brown_2004}. Consequently, parameter combinations which strongly (or weakly) influence model predictions can be revealed by identifying directions in parameter space which most (or least) influence model outputs. {\edit The analysis accounts for individual parameters acting together or against each other (compensating effects) and identifies sensitive and insensitive parameter combinations for the model-data fit \citep{gutenkunst_2007, brown_2003, Brown_2004, Gloria_2021_sloppy, Transtrum_2011,Transtrum_2014_MBAM}}. Hence, an analysis of model sloppiness could be used to identify and remove model mechanisms that only weakly impact predictions (or are weakly informed by the data), whilst accounting for hidden compensatory effects between individual model parameters. 

The concept of model sloppiness in model reduction methods has previously been explored by \cite{Transtrum_2014_MBAM}, who proposed the manifold boundary approximation method (MBAM). The MBAM uses an information theory-based approach whereby a parameter-independent geometric interpretation of the model is used to systematically reduce the effective degrees of freedom \citep{Transtrum_2014_MBAM}. {\edit However, alternative methods have been proposed for dimensionality reduction; for example, the active subspace method \citep{Constantine_2016_AS} similarly captures model-data fit sensitivities and can be used for individual parameter rankings through activity scores \citep{Constantine_2017_AS}.} More recently, \citet{Elevitch_2020} used a spectral analysis of the Hessian matrix, akin to the non-Bayesian analysis of model sloppiness \citep{Brown_2004}, to quantitatively rank parameter importance and thus determine which parameters should be estimated or fixed, rather than simplifying the model structure. 

In contrast to these methods, our approach is based solely on the analysis of a model's sloppiness.  Our method uses this analysis to identify insensitive groups of parameters which represent processes or mechanisms within a model, whilst accounting for the interdependencies between parameters. Thus, if model outputs are insensitive to changes in model parameters associated with a certain mechanism, such a mechanism is identified to have little effect on the overall model predictions (model-data fit). Using this sensitivity analysis, insensitive mechanisms are thus removed from the model (rather than being fixed or lumped) to produce a conceptually simpler model \citep{Transtrum_2014_MBAM} which maintains its predictive capability. As a result, the model remains expressed in terms of the parameters of interest and preserves mechanistic interpretability. Additionally, the method we propose can take advantage of Bayesian sensitivity matrices described recently elsewhere \citep{Gloria_2021_sloppy}. 

In this work, we showcase the potential for model sloppiness to inform strategic model reductions using a case study on a complex physiological model predicting coral calcification rates. We also use this case study to demonstrate that both Bayesian and non-Bayesian approaches to analysing model sloppiness \citep{Gloria_2021_sloppy,brown_2003} may be suitable for model reduction, although Bayesian approaches may be advantageous when the best-fit parameter values cannot be easily identified using standard search procedures (e.g.\ where the likelihood surface does not have a well defined peak).

\section{Methods}
\label{sec: Methods}

\subsection{Model calibration}
\label{Sec: Model calibration method}

Within complex models, there are often parameters which cannot be measured directly and must instead be estimated through a model-data calibration process. {\edit To analyse the model sloppiness, these unknown parameter values must first be estimated using either a classical (frequentist) calibration process to obtain a single-point estimate, or a Bayesian model-data calibration to obtain probabilistic distributions for parameters.}

\begin{figure}[H]
    \centering
    \includegraphics[width=\textwidth,trim={1cm 14cm 1cm 1cm}]{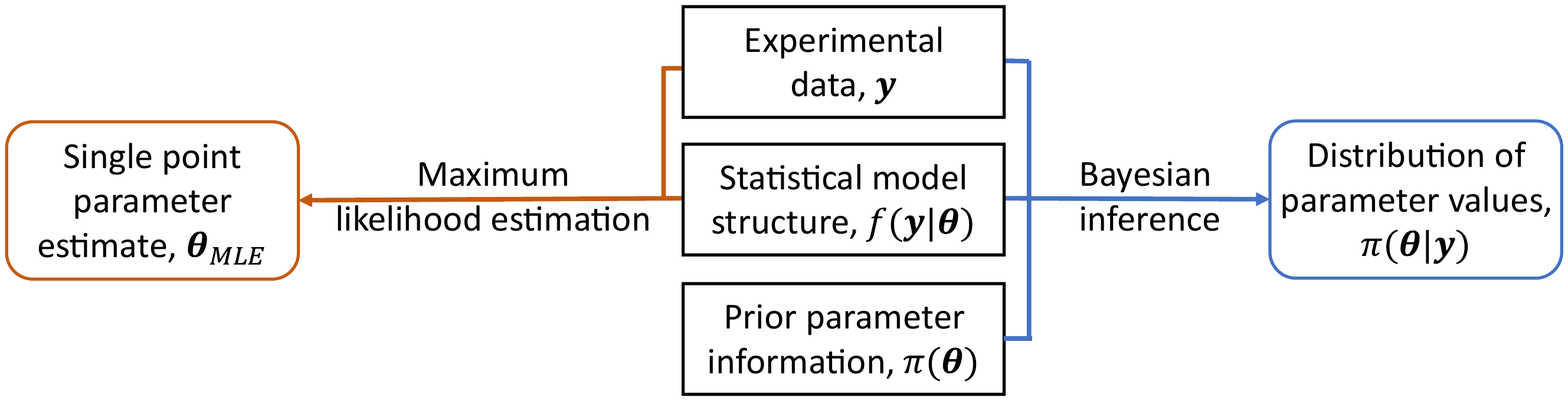}
    \caption{A conceptual figure, highlighting the difference between a model calibrated via MLE (left) and Bayesian inference (right). MLE combines experimental data with a model structure to produce a frequentist model. Bayesian inference also incorporates prior information about the model parameters to produce a probabilistic model that encapsulates parameter uncertainty.}
    \label{fig:Model calibration method}
\end{figure}

Single point estimates of unknown parameters can be obtained through maximum likelihood estimation (MLE), which is a common frequentist approach used for model calibration \citep{Jackson_2000}. More specifically, unknown model parameters $\bm{\theta}$ are estimated using observed data $\mathbf{y}$ by maximising an appropriately chosen likelihood function $f(\mathbf{y}|\bm{\theta})$. Under the common assumptions that measurement errors follow a Gaussian distribution of constant standard deviation, and that observations are independent, the likelihood that a set of data $\mathbf{y}$ is observed with a mean given by the model $\mathbf{y}_{\mathrm{model}}(\bm{\theta})$ and a standard deviation of $\sigma$ can be defined as
\begin{equation}
\label{eq:likelihood}
    f(\mathbf{y}|\bm{\theta}) = \prod _{i=1}^{n_d} f(y_{i}|\bm{\theta}) \\
    = \prod _{i=1}^{n_d} \frac{1}{\sqrt{2\pi} \sigma} \exp \left( -\frac{(y_{\mathrm{model},i}(\bm{\theta}) - y_{i})^2}{2\sigma^2} \right),
\end{equation}
where $\textbf{y}=\{y_1,y_2,\ldots,y_{n_d}\}$ is a dataset of $n_d$ independent observations, $y_{i}$ is the $i$th observation in the dataset, $f(\mathbf{y}|\bm{\theta})$ is the likelihood function, and $y_{\mathrm{model},i}(\bm{\theta})$ is the $i$th model prediction of the data given the conditions of observation $y_i$ and parameters $\bm{\theta}$. 
Formally, the MLE ($\bm{\theta}_{\mathrm{MLE}}$) is obtained by maximising the likelihood function, such that 
\begin{equation}
    \bm{\theta}_{\mathrm{MLE}} = \mathrm{argmax}_{\bm{\theta}}   f(\mathbf{y}|\bm{\theta}).
\end{equation}
Hence, $\bm{\theta}_{\mathrm{MLE}}$ represents the ``best-fit'' parameter values under the assumed statistical model structure and observed dataset. This process is represented as the left (orange) branch in Figure \ref{fig:Model calibration method}.  

However, this frequentist approach does not account for prior beliefs about the parameter values \citep{Uusitalo_2015}, and also ignores parameter values $\bm{\theta} \neq \bm{\theta}_{\mathrm{MLE}}$ which may also plausibly represent the data. To analyse the effect of the parameters on the behaviour of the system, the full range of values that a parameter could potentially take should be considered \citep{Jakeman_2006}. As an alternative to MLE, Bayesian inference can be used for model-data calibration to obtain a probability distribution for $\bm{\theta}$ that accounts for prior parameter information, an assumed model structure and the data \citep{Girolami_2008}. Such a distribution is called the posterior distribution as it represents a probability distribution for $\bm{\theta}$ \textit{after} considering the data. 

Using any known information about the parameters, a ``prior" distribution $\pi(\bm{\theta})$ is placed on $\bm{\theta}$. To obtain the posterior distribution, the prior distribution is multiplied by the likelihood function via Bayes' Theorem,
\begin{equation}
\label{eq:BayesRule} 
    \pi (\bm{\theta} |\mathbf{y}) \propto f(\mathbf{y}| \bm{\theta}) \pi(\bm{\theta}),
\end{equation}
where $\pi(\bm{\theta}|\mathbf{y})$ is the posterior distribution of the parameters, $f(\mathbf{y}|\bm{\theta})$ is the likelihood function and $\pi(\bm{\theta})$ is the prior distribution.
The posterior is a distribution of potential parameter values informed by the data and prior information such that it quantifies parametric uncertainty \citep{Girolami_2008}. 
In this paper, a sequential Monte Carlo (SMC) algorithm \citep{delmoral_2006,chopin_2002} was used to sample from the posterior distribution, providing a representative sample of plausible parameter values, given the available prior information and dataset (see \cite{drovandi_2011} or \cite{jeremiah_2012} for more information on SMC sampling). Bayesian inference for model calibration is depicted as the right (blue) branch of Figure \ref{fig:Model calibration method}.

\subsection{Analysing model sloppiness}
\label{Sec:model sloppiness method}

{\edit The analysis of model sloppiness is a type of sensitivity analysis which considers the model sensitivity to all parameters informed by a dataset \citep{Gloria_2021_sloppy}. As with global sensitivity analysis approaches \citep{Geris_2016,Marino_2008,Saltelli_1993}, the analysis of model sloppiness accounts for the model-data fit sensitivities of parameter \textit{combinations} across all parameters.} {\edit Thus, this approach can mathematically characterise} the parameter combinations that the model-data fit is most sensitive to  \citep{Transtrum_2015,brown_2003}. 

{\edit A sloppy model refers to a model where most of the model behaviour is captured through a few tightly constrained (stiff) parameter combinations (stiff eigenparameters) \citep{brown_2003}, which are highly influential on model predictions of the data \citep{Transtrum_2015}, but remains insensitive to many loosely constrained (sloppy) parameter combinations (sloppy eigenparameters).} Stiff and sloppy parameter combinations are identified through the eigendecomposition of a sensitivity matrix \citep{gutenkunst_2007}. {\edit While there are various approaches to the construction of the sensitivity matrix -- some of these are summarised in Section \ref{sec: sensitivity matrices} -- we focus here on how to obtain stiff/sloppy eigenparameters once this matrix is successfully computed.} 

{\edit Each eigenvector $\mathbf{v}_j$ of the sensitivity matrix indicates a key direction in parameter space that characterises the sensitivity of the model-data fit to changes in multiple parameters simultaneously. We can express each key direction in parameter space as a specified combination of parameter values -- known as an eigenparameter. Changing an eigenparameter's value is equivalent to moving the entire set of (original) model parameters along the direction of the eigenvector associated with that eigenparameter. As eigenvectors of the sensitivity matrix are by definition mutually orthogonal, if model parameters are logarithmically-transformed prior to calculation of the sensitivity matrix (which is a common practice when analysing sloppiness), each eigenparameter $\hat{\theta}_j$ of this matrix can be expressed as a linear combination of natural logarithms of model parameters, following \cite{Brown_2004},}
\begin{align}
    \log \hat{\theta}_j &= {v_{j,1}}\log{\theta_1}+{v_{j,2}}\log{\theta_2}+\cdots+{v_{j,n_p}}\log{\theta_{n_p}}, \nonumber \\
    \label{Eq:Eigenparameter}
    \therefore \hat{\theta}_j &= {\theta_1}^{v_{j,1}}{\theta_2}^{v_{j,2}}\cdots{\theta_{n_p}}^{v_{j,{n_p}}} , 
\end{align} 
where $\mathbf{v}_j = [v_{j,1}, v_{j,2}, ..., v_{j,{n_p}}]$ is the $j$th eigenvector of the sensitivity matrix,  ${n_p}$ is the number of model parameters, and $\theta_i$ is the $i$th parameter in the model. It should be noted that expressing eigenparameters $\hat{\theta}_j$ as a product of all model parameters to the power of the different exponents as shown in Equation \eqref{Eq:Eigenparameter} is only possible if the parameters are scaled by their logarithm when estimating the sensitivity matrix. {\edit Additionally, standard renormalisations can be applied such that the exponents $v_{j,i}$ in Equation \eqref{Eq:Eigenparameter} are rescaled to be between $-1$ and $1$. }

{\edit In Equation \eqref{Eq:Eigenparameter}, each eigenparameter $\hat{\theta}_j$ has a corresponding eigenvalue $\lambda_j$ which attributes a magnitude to the direction of the eigenparameter. The largest eigenvalue ($\lambda_1$) corresponds to the stiffest (most sensitive) eigenparameter, and the smallest eigenvalue ($\lambda_{n_p}$) corresponds to the sloppiest (least sensitive) eigenparameter \citep{Transtrum_2015}. Therefore, comparing the eigenvalues of all eigenparameters indicates the relative impact that each parameter combination $\hat{\theta}_j$ has on the model-data fit. In our implementation of this approach, we scale all eigenvalues $\lambda_j$ by dividing them by the largest eigenvalue $\lambda_1$ to indicate the relative importance of the eigenparameters. Thus, $0 < \lambda_j /\lambda_1 \leq 1$ for all eigenparameters $\lambda_j, j=1,\ldots,n_p$. }

{\edit The sloppy analysis results in a list of $n_p$ weighted parameter combinations $\hat{\theta}_j$ ranked by their influence on the model-data fit via the scaled eigenvalues $\lambda_j /\lambda_1$. This process is summarised in Algorithm \ref{Alg: Pseudo-code for analysing model sloppiness} (see also \cite{Gloria_2021_sloppy} for an overview of model sloppiness). }

{\edit 
\begin{algorithm}[H]
\begin{enumerate}
    \item Perform model calibration to obtain estimate(s) of parameters $\bm{\theta}$.
    \item Calculate sensitivity matrix $S$ with respect to the log-parameters, $\phi_i=\log \theta_i$.
    \item Find all eigenvalues ${\lambda_j}$ and eigenvectors $\bm{v}_j, j=1,\ldots,n_p$ of the sensitivity matrix $S$.
    \item Order the eigenvectors by their influence using their associated eigenvalues ${\lambda}_j$, such that $\lambda_1 \geq \lambda_2 \geq \cdots \geq  \lambda_{n_p}$.
    \item Calculate the rescaled eigenvalues $\lambda_j/\lambda_1$, for all $j=1,\ldots,n_p$.
    \item Renormalise each eigenvector $\bm{v}_j \leftarrow \bm{v}_j/\mathrm{max} (|\bm{v}_j|) $.
    \item Report the eigenparameters using Equation \eqref{Eq:Eigenparameter} and the renormalised eigenvectors obtained from Step 6, ordered in importance by the magnitude of the corresponding eigenvalues. 
\end{enumerate}
    \caption{Process used for analysing the model sloppiness}
\label{Alg: Pseudo-code for analysing model sloppiness}
\end{algorithm}
}

\subsection{Sensitivity matrices}
\label{sec: sensitivity matrices}

{\edit The key quantity required for analysing sloppiness of a model fitted to data is the sensitivity matrix $S$. This is a square symmetric matrix of size $n_p \times n_p$, in which $n_p$ is the number of estimated model parameters, excluding parameters that represent measurement error (e.g. standard deviation $\sigma$ in Equation \eqref{eq:likelihood}) as the latter can yield eigenparameters that are trivial or difficult to interpret \citep{Gloria_2021_sloppy}.} There are various approaches to calculate the sensitivity matrix (see \cite{Gloria_2021_sloppy} for an overview). {\edit Although based on similar dimension reduction techniques (e.g.,\ posterior covariance, likelihood informed subspace, or the active subspace methods), each sensitivity matrix} considers different sources of information for the model parameters and their interdependencies. For example, the chosen sensitivity matrix may acknowledge or aim to exclude prior beliefs about the model parameters when identifying key parameter interdependencies (e.g.\ the posterior covariance method and likelihood informed subspace methods, respectively). 

{\edit In this paper three sensitivity matrices (Table \ref{tab: Sloppy approaches}) were used to explore model reduction. Firstly, the Hessian matrix of the log-likelihood is used to capture the model-data fit around one point in parameter space based on the likelihood surface \citep{brown_2003}. Secondly, the posterior covariance is a variance-based method which looks at model-data fit sensitivities over the full posterior parameter-space \citep{brown_2003,Gloria_2021_sloppy}. Thirdly, the likelihood-informed subspace (LIS) method captures the model-data sensitivities of the posterior sample relative to the prior distribution \citep{Cui_2014}, so it can be used in comparison with the posterior covariance method to identify the informativeness of the prior distribution \citep{Gloria_2021_sloppy}. We chose to use these three sensitivity matrices because the interpretation of the analysis of sloppiness for these matrices can provide uniquely different information, as documented recently for multiple simulation problems \citep{Gloria_2021_sloppy}. However, there are other sensitivity matrices that could also be used, such as the Levenberg-Marquardt Hessian \citep{brown_2003,gutenkunst_2007}, the matrix arising from the active subspace method \citep{Constantine_2016_AS}, or a likelihood-free approximation of LIS \citep{Beskos_2018}. } 

Sensitivity matrices are often calculated using logarithmically-scaled parameter values \citep{Brown_2004}. This rescaling enforces positivity constraints on the model parameters, helps prevent scaling issues between parameters with units possessing different orders of magnitude, and allows each eigenparameter to be expressed as a product rather than as a sum, as in Equation \eqref{Eq:Eigenparameter} \citep{Gloria_2021_sloppy}. Here, we denote the logarithmically-rescaled parameters as $\phi_i=\log \theta_i$, for all $i=1,\ldots,{n_p}$ parameters {\edit (see Step 2 of Algorithm \ref{Alg: Pseudo-code for analysing model sloppiness})}. 

\begin{table}[H]
    \centering
    \small 
    \begin{tabular}{m{0.11\linewidth}|c|m{0.27\linewidth}|m{0.22\linewidth}|m{0.1\linewidth}}
         \textbf{Sensitivity Matrix} & \textbf{Bayesian} & \textbf{Formula} & \textbf{Features} & \textbf{Reference} \\\hline
         Hessian evaluated at the MLE $(S_H)$ & No &  \begin{equation*} \abovedisplayskip=0pt H_{i,j}=\frac{\partial^2 \log f}{\partial\phi_i\partial\phi_j}|_{\phi=\hat{\phi}_{\mathrm{MLE}}}, \end{equation*} where $H_{ij}$ is the $(i,j)$th element of Hessian matrix ($S_H$) evaluated at $\hat{\phi}_{\mathrm{MLE}}$, and $f$ is the likelihood function. & Looks at curvature of model-data fit surface at the MLE. & \citet{brown_2003} \\ \hline
         Posterior covariance method $(S_P)$ & Yes & \begin{equation*} \abovedisplayskip=0pt S_{P} = \Sigma^{-1}, \end{equation*} where $\Sigma$ is the empirical covariance matrix of $\{\phi_m\}_{m=1}^M$, the set of $M$ posterior samples. & Uses the posterior sample to estimate model-data sensitivities over a range of parameter values. & \citet{brown_2003,Gloria_2021_sloppy} \\ \hline
         Likelihood informed subspace method $(S_L)$ & Yes & \begin{equation*} \abovedisplayskip=0pt S_{L} \approx \frac{1}{M} \sum_{m=1}^M L^T H(\phi_m) L, \end{equation*} where $\{ \phi_m\}_{m=1}^M$ is the set of $M$ posterior samples, $H(\phi_m)$ is the Hessian matrix evaluated at $\phi_m$, and $L$ is the Cholesky factor of the covariance matrix $\Omega$ of the prior distribution $\pi(\phi)$, such that $\Omega = LL^T$. & Estimates model-data fit sensitivities by comparing where the data is most informative relative to the prior distribution. Can identify how informative the prior distribution is on the posterior distribution, when compared to the posterior covariance method. & \citet{Gloria_2021_sloppy,Cui_2014}
    \end{tabular}
    \caption{Three methods of constructing the sensitivity matrix for an analysis of model sloppiness. Each sensitivity matrix captures different features of the model-data fit. For the Bayesian sensitivity matrices, it is assumed here that the posterior distribution is approximated by a sufficiently large number $M$ of equally weighted posterior samples.}
    \label{tab: Sloppy approaches}
\end{table}

\subsection{Model reduction}
\label{sec: model reduction}

Model reduction aims to simplify the model whilst minimising the loss of predictive power of the model \citep{Snowden_2017,Jeong_2018}. Given that model sloppiness can identify sensitive parameter combinations, we propose that model reduction can be informed by considering the removal of mechanisms that contribute negligibly to these sensitive parameter combinations. {\edit This approach is similar, although not equivalent, to factor-fixing of sets of parameters from a variance-based sensitivity analysis \citep{Saltelli_2004}}. 

Previous work has shown that analysing model sloppiness can reduce the number of model parameters via iteratively removing the sloppiest eigenparameter and simultaneously adapting the model with limiting approximations \citep{Transtrum_2014_MBAM}. Here, we instead investigate the possibility that, if one or more parameters which together characterise an entire process or mechanism within the model have little contribution to the stiffest eigenparameters, this process or mechanism may contribute very little to the model's ability to predict the data that was used for its calibration. Hence, this analysis can identify which \textit{mechanisms} the model-data fit is insensitive to {\edit-- similar to variance-based sensitivity analyses, such as Sobol's indices \citep{Sobol_1993}, which can analyse sets of parameters or model structures \citep{Mara_2017}. These identified mechanisms can be potentially removed} from the model with little effect on the model's predictive power. 

\begin{figure}[H]
    \centering
    \includegraphics[width=12cm,trim={1cm 16cm 1cm 1cm}]{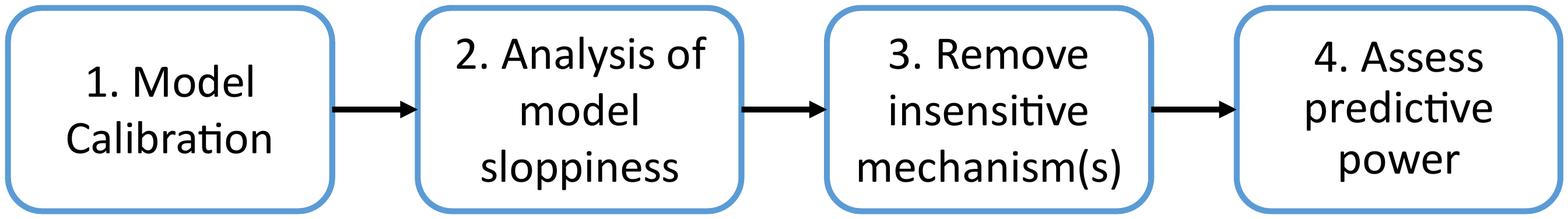}
    \caption{A conceptual diagram of the model reduction process informed by the analysis of model sloppiness. Firstly, the model is calibrated (Section \ref{Sec: Model calibration method}) using the Bayesian and/or non-Bayesian methods depicted in Figure \ref{fig:Model calibration method}. The second step is an analysis of model sloppiness (Section \ref{Sec:model sloppiness method}) to identify model mechanisms which weakly inform model predictions. Thirdly, the insensitive mechanisms identified in the analysis of model sloppiness are removed from the model and the simplified models are calibrated to the data. Lastly, the predictive power of the reduced model(s) can be assessed, e.g.\ via goodness-of-fit and/or model selection metrics, to determine the best model for the application.}
    \label{fig:Model reduction process}
\end{figure}

We suggest that a general quantitative method for selecting the most insensitive model mechanism to be removed would not be appropriate for all models. Instead, the analysis of model sloppiness is used as a source of information for guiding model reductions, paired with an understanding of the model and data being considered. {\edit We also recommend quantitatively comparing the original and reduced models using tools such as model evidence, Bayes factors, the Akaike information criterion or the Bayesian information criterion \citep{Tredennick_2021,Kass_1995_BF}}.  The full process of model reduction informed by the analysis of model sloppiness that we propose and investigate here is depicted in Figure \ref{fig:Model reduction process}.

\section{Case study}
\label{sec: Results}

\subsection{Coral calcification model}

To demonstrate using the analysis of model sloppiness as a method for strategic model reduction, this method was applied to a data-calibrated model predicting coral calcification rates. Calcification rates are a common metric for measuring the health of coral reefs \citep{Erez_2011}. The process of coral calcification involves the coral laying down its skeleton, resulting in the spatial extension of coral reef structures \citep{andersson_2013}. This process is vital for the entire ecosystem, because it forms a habitat for a diverse range of marine life as well as providing a stuctural framework for barrier reefs \citep{hoegh_2017}.

\begin{figure}  [H]
    \centering
    \includegraphics[width=\textwidth]{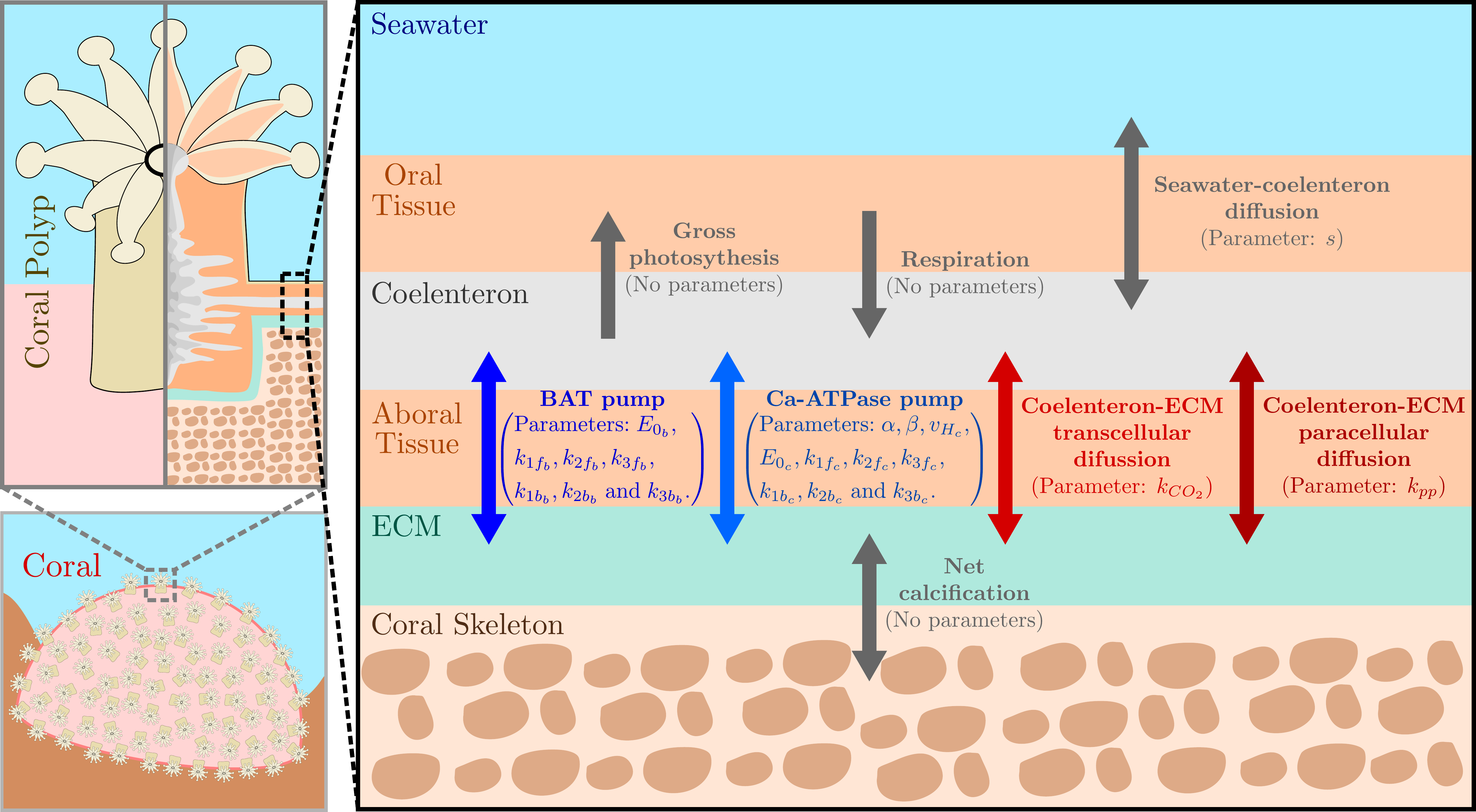}
    \caption{Conceptual model of the chemical processes and reactions within a coral polyp, proposed by \cite{Galli_2018}. Each of the arrows in this diagram represents a mechanism which affects the chemical species concentrations in the model. This model predicts the calcification reaction rate in the extracellular calcifying medium (ECM) based on environmental conditions.}
    \label{fig:Conceptual model}
\end{figure}

The model and data used for this case study are reported by \cite{Galli_2018}; this is the most recent and comprehensive model for coral reef calcification rates. The model prediction of calcification rates are obtained from the steady state solution of a system of nonlinear ordinary differential equations (ODEs), which together simulate the chemical composition of relevant molecules and ions throughout various physiological compartments of a coral polyp (Figure \ref{fig:Conceptual model}). {\edit There are eight mechanisms within the model which represent different chemical processes and reactions through coacting and counteracting flux terms (Table \ref{tab: Mechanisms})}, hence, some of these mechanisms could potentially be removed. Further description of the model is provided in Supplementary Material, Section S.1. \cite{Galli_2018} calibrated the 20 unknown parameters of the model (see Supplementary Material Section S.2 for details) to an experimental dataset containing 16 data-points obtained from \cite{rodolfo_2010}. Each data point consists of paired measurements of calcification rates and the environmental conditions under which this rate was measured (data shown in Table 1 of \citeauthor{Galli_2018}, \citeyear{Galli_2018}). 

\begin{table}[H]
    \centering
    \small 
    \edit
    \begin{tabular}{p{0.22\linewidth}|p{0.45\linewidth}|p{0.24\linewidth}}
        \textbf{Mechanism} & \textbf{Description} & \textbf{Parameters} \\ \hline
        Net calcification & The net effect of calcification and dissolution chemical reactions which produce and remove coral skeleton respectively & N/A \\ \hline
        Gross photosynthesis & A chemical reaction caused by zooxanthellae algae yielding carbon reduction in the coelenteron & N/A\\ \hline
        Respiration & A chemical reaction caused by zooxanthellae algae yielding carbon gain in the coelenteron & N/A\\ \hline
        Seawater-coelenteron diffusion & Diffusion for all chemical species between seawater and coelenteron & $s$\\ \hline
        Coelenteron-ECM paracellular diffusion & Paracellular diffusion for all chemical species between the coelenteron and ECM & $k_{pp}$\\ \hline
        Coelenteron-ECM transcellular diffusion & Transcellular diffusion of carbon dioxide between the coelenteron and ECM & $k_{CO2}$\\ \hline
        Ca-ATPase pump & The active transport of calcium ions through the aboral tissue, driven by ATP & $\alpha$, $\beta$, $v_{H_c}$, $E0_c$, $k_{1f_c}$, $k_{2f_c}$, $k_{3f_c}$, $k_{1b_c}$, $k_{2b_c}$, $k_{3b_c}$ \\ \hline
        BAT pump & The active transport of bicarbonate anions through the aboral tissue & $E0_b$, $k_{1f_b}$, $k_{2f_b}$, $k_{3f_b}$, $k_{1b_b}$, $k_{2b_b}$, $k_{3b_b}$\\ 
    \end{tabular}
    \caption{The eight model mechanisms and their associated parameters in the model proposed by \cite{Galli_2018}. Note that the first four mechanisms cannot be removed because they are the rate of interest we wish to output (net calcification), data inputs with no associated parameters (gross photosynthesis and respiration), and a mechanism critical for the model's original purpose of  simulating coral responses to ocean acidification (seawater-coelenteron diffusion).}
    \label{tab: Mechanisms}
\end{table}

\subsection{Model calibration of the coral calcification model}
\label{sec: model calibration results}
The coral calcification model was calibrated to the data (see Table 1 of \citeauthor{Galli_2018}, \citeyear{Galli_2018}) using both MLE and Bayesian inference methods. The likelihood function was defined based on Gaussian errors as in Equation \eqref{eq:likelihood}, where the data was assumed to be normally distributed with a mean calcification rate according to the model proposed by \citet{Galli_2018}, and a constant and unknown standard deviation of $\sigma$. 

\begin{figure}[H]
    \centering
    \includegraphics[width=0.85\textwidth]{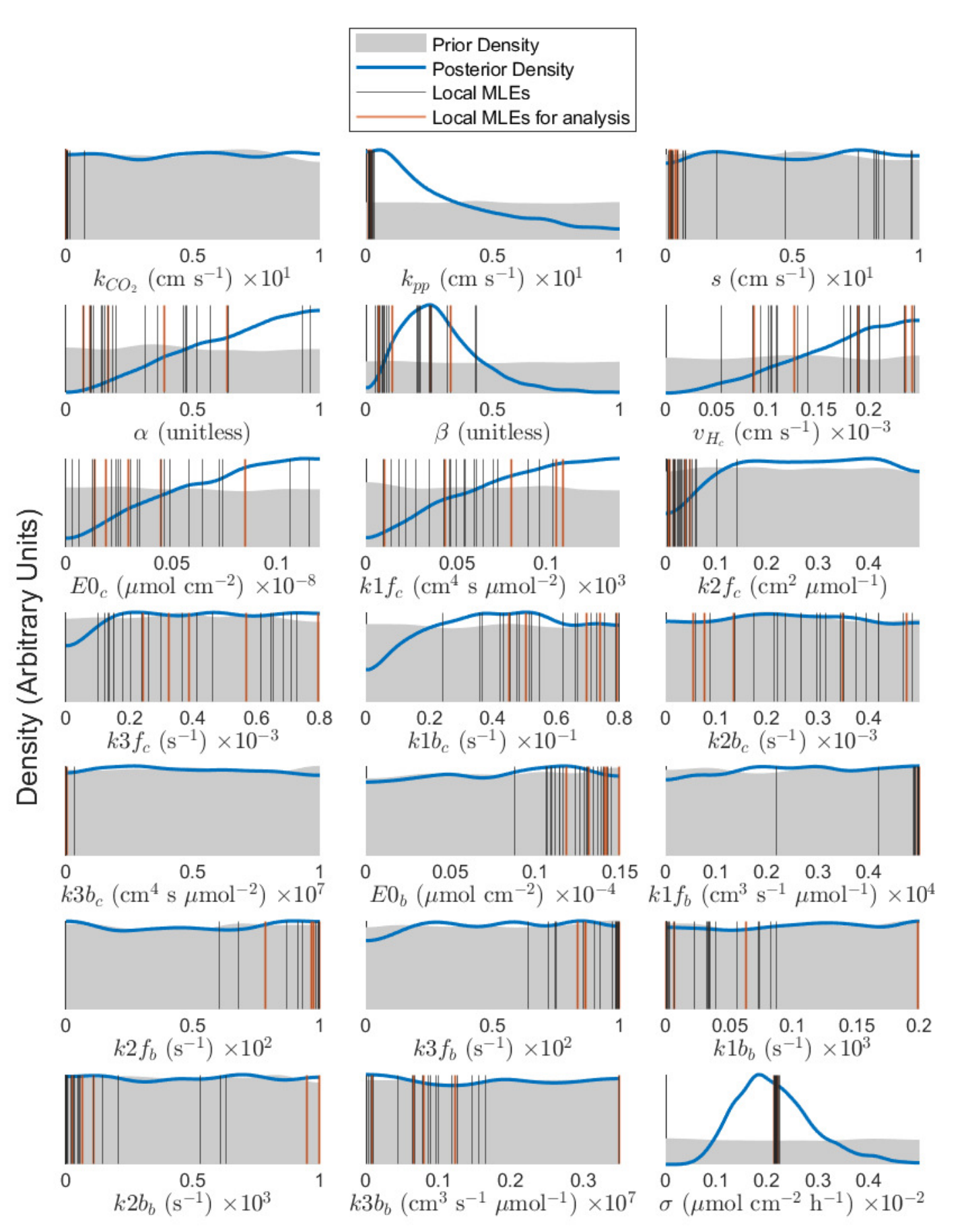}
    \caption{Parameter estimates obtained through both Bayesian inference and MLE model-data calibration methods. The prior and posterior distributions obtained via Bayesian inference are indicated as the grey areas and thick blue lines respectively, showing the change to the distribution as informed by the data. The vertical lines indicate different local likelihood maxima with similarly high likelihoods, with those in orange possessing the highest likelihood function values and thus kept for further analysis. Notice that the local likelihood maxima are distinct and cover a vast range of the parameter space, indicating the complexity of the high-dimensional likelihood surface.}
    \label{fig:LMLE marginals}
\end{figure}

In cases where the model-data fit surface is complex, local optima can misguide optimisation algorithms such that they do not converge on the global MLE \citep{Transtrum_2011}. In the present case study, the likelihood function was high-dimensional due to its dependence on 20 model parameters and had no clearly defined peak, hence the global MLE was difficult to obtain. Instead, 100 local MLEs were obtained and the five with the highest likelihood values were used in further analysis. These local maxima were identified using a gradient-based non-linear function minimisation tool (Matlab's \textit{fmincon} function, described in \citeauthor{fmincon_matlab}, \citeyear{fmincon_matlab}) using various initial search locations. Figure \ref{fig:LMLE marginals} shows 25 of the identified local maxima with the highest likelihood values (black vertical lines), which are spread across many parameter values, indicating the complexity of the high-dimensional likelihood surface. These distinct parameter estimates each have similarly high likelihoods and this goodness-of-fit is reflected by their predictions of calcification rates when compared to the observations (black asterisks in Figure S1 of Supplementary Material Section S.3). {\edit Only one point-estimate (the MLE) is needed for the analysis of model sloppiness if the likelihood surface has a well-defined peak. However, when the likelihood surface is multi-modal and/or flat in certain directions (as is common in sloppy models) it can be difficult to identify the global MLE.  Thus, to analyse the reproducibility of the results, we evaluated the Hessian matrix at five different sets of parameter values -- those that yielded the highest values of the likelihood function (orange vertical lines in Figure \ref{fig:LMLE marginals}, and orange asterisks in Figure S1 of Supplementary Material Section S.3) algorithm. }

To calibrate the coral calcification model a second time, using Bayesian inference instead, we specified uniform prior distributions (grey shaded regions in Figure \ref{fig:LMLE marginals}) as most of the parameters for this model were only known to be strictly positive. The posterior distribution was approximated using an adaptive SMC algorithm \citep{drovandi_2011} and this algorithm was run {\edit multiple} times independently to test the reproducibility of both the posterior sample and the later analyses of model sloppiness {\edit (we used five independent posterior samples)}. {\edit The estimated marginal distributions were visually indistinguishable for the independently produced posterior samples, indicating that the posterior sample was highly reproducible for a posterior sample size of 5000.} Further details of the model calibration procedure performed using Bayesian inference are provided in Supplementary Material, Section S.4. The estimated posterior marginal densities obtained for the model parameters (Figure \ref{fig:LMLE marginals}) revealed large uncertainty in many of the model parameters, with only two parameters strongly informed by the data ($k_{pp}$ and $\beta$). This result is expected given the limited size of the dataset available for model-data calibration.

\subsection{Analysing sloppiness of the coral calcification model}
\label{sec: Sloppiness results}

After the model-data calibration, we used the analysis of model sloppiness to unravel strong parameter interdependencies. This analysis was conducted for each of the three sensitivity matrices listed in Table \ref{tab: Sloppy approaches}: the Hessian approach evaluated at each of the five local MLEs, the posterior covariance method, and the LIS approach, with both of the latter evaluated from the posterior distribution samples obtained from the SMC algorithm.   

\begin{figure}[H]
    \centering
\includegraphics[width=\textwidth]{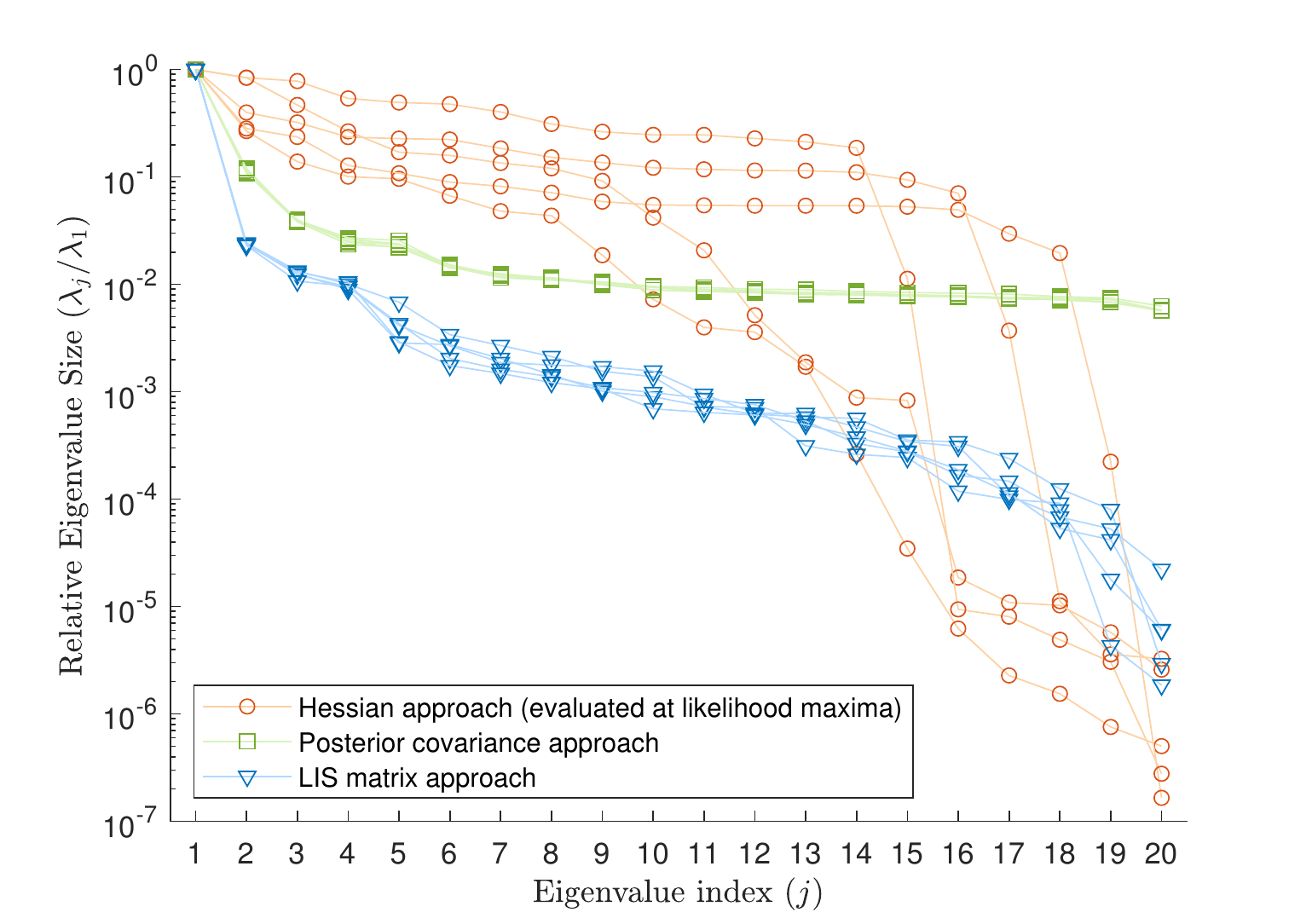}
    \caption{Relative eigenvalue sizes ($\lambda_j/\lambda_1$) compared to the largest eigenvalue ($\lambda_1$) for each of the approaches to analysing model sloppiness. Each of the MLE Hessian analyses of model sloppiness were based on parameter estimates (orange vertical lines in Figure \ref{fig:LMLE marginals}) from different local maxima. Both the posterior covariance method and LIS sensitivity matrix analyses were conducted on five independent sets of posterior samples obtained using Bayesian inference. Notice that the posterior covariance method leads to extremely similar eigenvalue spectra for the five independent sets of posterior samples.}
    \label{fig:relative importance}
\end{figure}

Figure \ref{fig:relative importance} shows the size of the eigenvalues relative to the largest, after eigendecomposition of each sensitivity matrix. Here, the MLE Hessian approach (Table \ref{tab: Sloppy approaches}, first row) leads to an inconsistent decay in eigenvalue spectra for the five different local MLEs used to evaluate the  sensitivity matrix. This result is unsurprising, as the parameter values of each local MLE are in distinctly different locations of the likelihood surface, so the  sensitivity of the model-data fit changes in these different local parameter spaces considered. In contrast, the posterior covariance method (Table \ref{tab: Sloppy approaches}, second row) produces a consistent decay in eigenvalue size across independent posterior samples. The consistent decay is likely because the full range of parameter values from the posterior are considered when analysing the model-data sensitivity, achieving a global analysis of the posterior surface. Finally, the decline in relative eigenvalue size is much more rapid for the LIS framework (Table \ref{tab: Sloppy approaches}, third row, and Figure \ref{fig:relative importance}, blue triangles), when compared to the posterior covariance approach. Here, the relative eigenvalue size for the fifth LIS eigenparameter is comparable to the twentieth (i.e.\ smallest eigenvalue) from the posterior covariance method.

\subsubsection{Hessian matrix evaluated at local likelihood maxima}
First, the non-Bayesian approach to analysis of sloppiness (i.e.\ using the Hessian matrix) was considered. As the Hessian matrix could not be analytically computed for the model in this case study, the finite differences method \citep{beers_2007_finitediffs} was used to numerically approximate the Hessian matrix. {\edit When using finite differences, the likelihood function is evaluated very close to the MLE, using a step-size $\Delta \theta_i = \delta \times \theta_i, i=1,\ldots, n_p$, which is as small as possible to minimize truncation error. However, a step-size too small will result in numerical issues (round-off errors). For this application, we used $\delta = 10^{-2}$, as larger step-sizes yielded the same results, but smaller step-sizes yielded inconsistent results which we attributed to numerical errors.}  The Hessian matrix was evaluated at five different parameter sets -- those that yielded the highest values of the likelihood function (orange vertical lines in Figure \ref{fig:LMLE marginals}).

\begin{figure}[H]
    \centering
    \includegraphics[width=\textwidth,trim={0cm 8.5cm 5.0cm 0cm},clip]{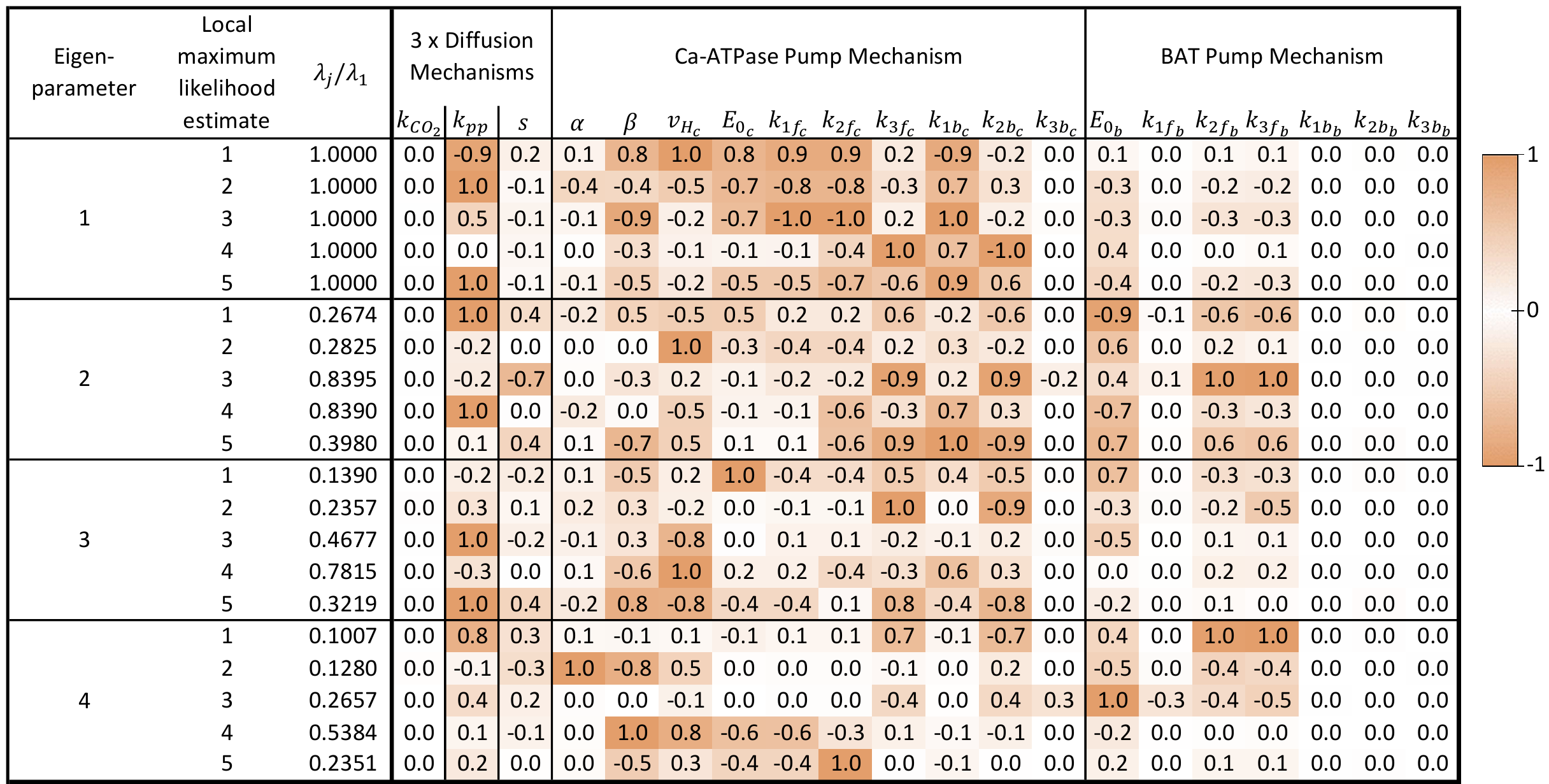}
    \caption{Eigenvector element values for eigenparameters identified using the MLE Hessian approach to an analysis of sloppiness. These eigenparameters correspond to the four largest eigenvalues, and so are ordered from highest relative importance to lowest. For each eigenparameter, the results of five high likelihood parameter estimates are compared to indicate the consistency of results. {\edit Here we only consider four of twenty eigenparameters to show the inconsistency between results based on different local MLEs in the four most sensitive parameter combinations.} For each eigenparameter, the values were normalised by the leading eigenvector value, such that they are rescaled to be between -1 and 1 inclusive. Here the colour darkens as the absolute values of the eigenvector values increase from 0 to 1, such that the model-data fit is more sensitive to darker eigenvector values. Notice that each of the parameters has been grouped based on its mechanistic function in the model (Figure \ref{fig:Conceptual model}). Additionally, the relative size of each eigenvalue when compared to the leading eigenvalue for each sample has been included in the column $\lambda_j/\lambda_1$. }   
    \label{fig:MLE original}
\end{figure}

After eigendecomposition of the Hessian matrix, evaluated at each parameter set, we used Equation (\ref{Eq:Eigenparameter}) to identify parameter interdependencies (eigenparameters). Figure \ref{fig:MLE original} shows the eigenparameters that correspond to the four largest eigenvalues ($\lambda_1, \lambda_2, \lambda_3$ and $\lambda_4$ for the orange circles in Figure \ref{fig:relative importance}), for each of the five likelihood maxima considered. In each row of the matrix depicted in Figure \ref{fig:MLE original}, the relative contribution of a given ($i$th) parameter to the ($j$th) eigenparameter (parameter combination) is indicated by the listed value. This value mathematically represents the eigenvector element value $v_{j,i}$ and can be interpreted as the magnitude of the $i$th parameter's exponent in the expression (Equation (\ref{Eq:Eigenparameter})) for the $j$th eigenparameter. Standard renormalisations during eigendecomposition ensure that $-1\leq v_{j,i}\leq 1,\ \forall i,j$, such that exponents that are close to $1$ or $-1$ indicate strong contribution of the parameter to the eigenparameter, and exponents close to $0$ indicate negligible contribution of the parameter to the eigenparameter. For example, in the first row of Figure \ref{fig:MLE original}, the parameter $k_{pp}$ has an exponent of $-0.9$, so strongly contributes to the first (stiffest) eigenparameter. In contrast, the parameter $k_{CO2}$ in the first row contributes negligibly to the stiffest eigenparameter.

The parameters that contribute to the stiffest eigenparameter are largely associated with the coelenteron-ECM paracellular diffusion mechanism (represented as $k_{pp}$) and the Ca-ATPase pump mechanism (Figure \ref{fig:MLE original}). In contrast, parameters associated with the coelenteron-ECM transcellular diffusion mechanism (represented as $k_{CO2}$), seawater-coelenteron diffusion mechanism (represented as $s$) and the BAT pump mechanism all contribute to a lesser extent to the stiffest eigenparameter. While this result is generally true when considering different point estimates (local MLEs), the relative contribution of each individual parameter to the eigenparameters depends on the local MLE used to evaluate the Hessian matrix (Figure \ref{fig:MLE original}). As each of the likelihood maxima had distinct parameter values, this indicates that the sensitivity of the model-data fit to parameter combinations depends on the local parameter space considered. This motivates the need for a sensitivity matrix which captures the model-data fit sensitivity across a range of potential parameter values, such as the posterior covariance method.

\subsubsection{Posterior covariance method}
\label{sec: results PCA}
\begin{figure}[H]
    \centering
    \includegraphics[width = \textwidth,trim={0cm, 9.0cm, 5.7cm, 0cm},clip]{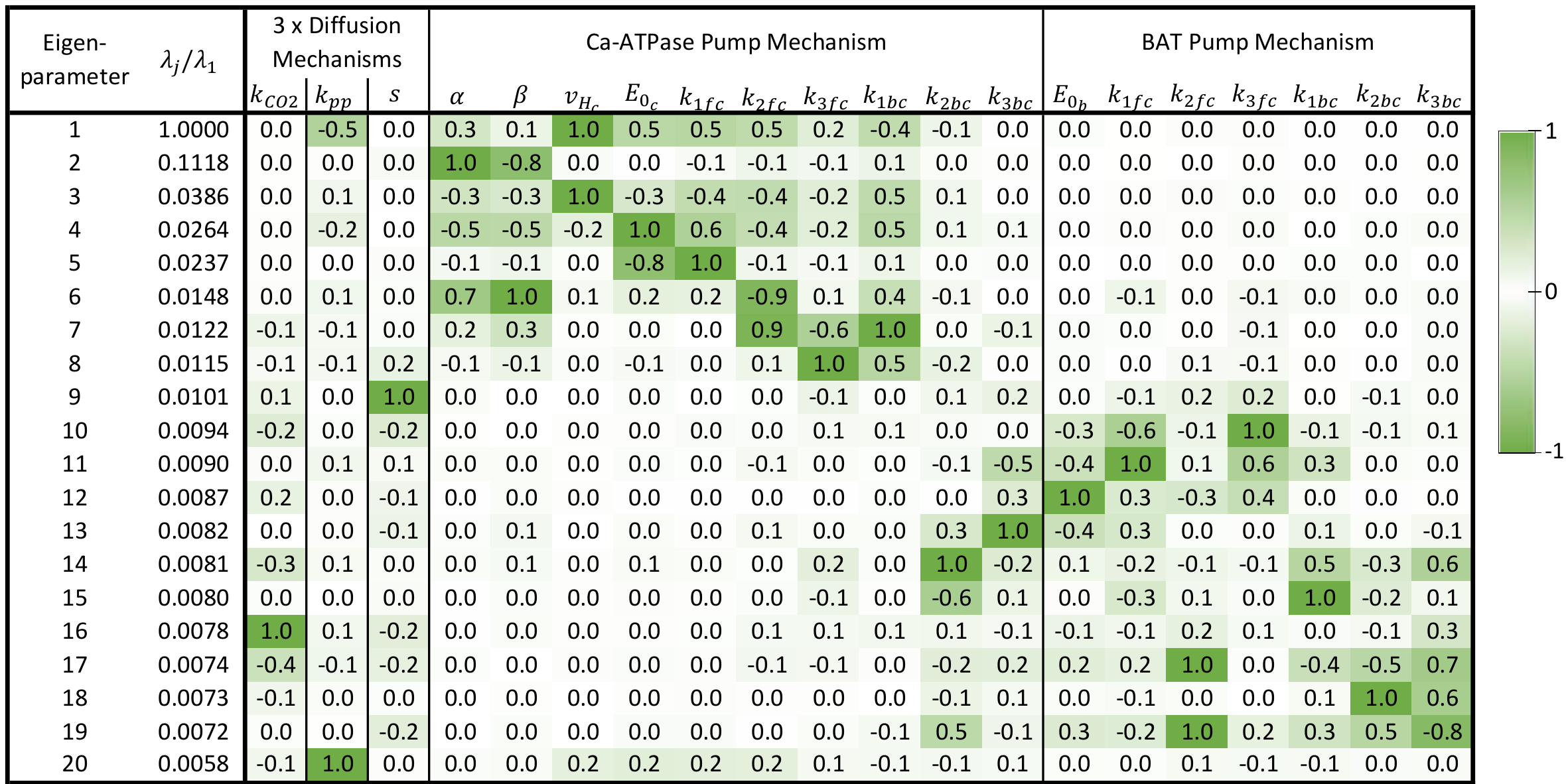}
    \caption{Eigenvector values for eigenparameters identified using a posterior covariance analysis of sloppiness, based on one set of posterior samples. {\edit Here we show the full analysis of model sloppiness, i.e.\ all 20 eigenparameters. (See Supplementary Material Section S.5 for confirmation of the consistency across different sets of posterior samples.)} See the caption of Figure \ref{fig:MLE original} for further interpretation of this figure.}
    \label{fig:PCA original}
\end{figure}

As an alternative to the MLE Hessian approach, the posterior covariance method considers a sample of the posterior distribution to obtain the key eigendirections (parameter interdependencies). For the analysis performed on one representative set of posterior samples (Figure \ref{fig:PCA original}), the two stiffest eigenparameters indicate that the coelenteron-ECM paracellular diffusion (represented as $k_{pp}$), and Ca-ATPase pump mechanisms strongly inform model predictions, in agreement with the results obtained for the Hessian matrix. This suggests that within this dataset, the calcification rate was strongly informed by the balance of these mechanisms within the coral polyp. 

Unlike the Hessian-based results, the parameters associated with the BAT pump, seawater-coelenteron diffusion mechanism (represented as $s$) and the coelenteron-ECM transcellular diffusion (represented as $k_{CO2}$) mechanisms contribute little to the model behaviour within the seven stiffest eigenparameters. Given the decay in relative eigenvalue size beyond the seventh eigenparameter (more than two orders of magnitude, see Figure \ref{fig:relative importance}), the analysis suggests that these latter mechanisms may not be necessary to maintain a good model-data fit for  calcification rate predictions.

In our case study, the posterior covariance method yielded substantially more consistent results across different sets of posterior samples (Supplementary Material Section S.5) when compared to the MLE Hessian approach across different local MLEs (Figure \ref{fig:MLE original}). When using a posterior covariance method, the analysis yielded consistent eigenvector values for the stiffest nine eigenparameters in Figure \ref{fig:PCA original} across independently generated sets of posterior samples (Figure S2  of Supplementary Material Section S.5). Here, differences amongst eigenparameters having small eigenvalues are expected as the model-data fit is weakly sensitive to sloppy eigendirections \citep{brown_2003}.

\subsubsection{Likelihood informed subspace sensitivity matrix}
Lastly, the analysis of model sloppiness using a LIS method was considered. One advantage of using both a posterior covariance and LIS approach to analysing sloppiness is that the influence of the prior distribution on the posterior sample can be identified \citep{Gloria_2021_sloppy}. More specifically, if there are substantial differences in the directions of stiff eigenparameters between the posterior covariance and LIS methods, this indicates that the prior is having a large impact on the stiff eigenparameter directions found using posterior covariance, and therefore the prior itself is strongly informing the posterior. For calculation of the LIS matrix, finite differences was also used to approximate the Hessian, with the step-size $\delta = 10^{-2}$. 

\begin{figure}[H]
    \centering
    \includegraphics[width=\textwidth,trim={0cm 15.9cm 4.5cm 0cm},clip]{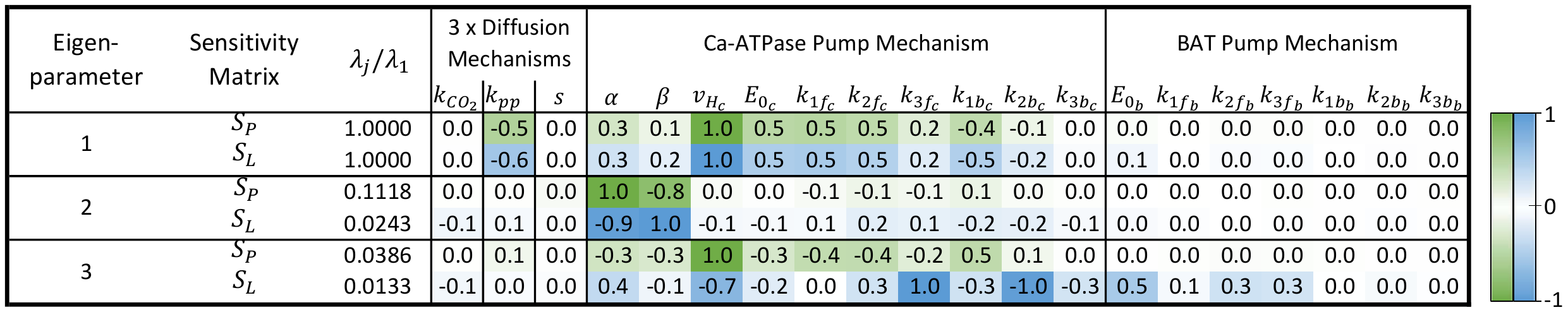}
    \caption{Comparison of eigenvector values of the three stiffest eigenparameters of the analysis of model sloppiness from LIS ($S_L$) and posterior covariance ($S_P$) sensitivity matrices (blue and green respectively). {\edit Here we only consider three of twenty eigenparameters to compare the consistency between approaches for the three most sensitive parameter combinations.} See the caption of Figure \ref{fig:MLE original} for further interpretation of this figure.}
    \label{fig:LIS original}
\end{figure}

Looking at the two stiffest eigenparameters (Figure \ref{fig:LIS original}), the LIS analysis of sloppiness gives a similar result to that of the posterior covariance method {\edit -- the relative magnitude of the elements within the two eigenvectors are very similar between methods.} However, beyond the second eigenparameter, the results of the two approaches begin to differ and the LIS results become inconsistent across independent posterior samples (Supplementary Material Section S.6). Here we note that the relative decay in eigenvalue size is much more rapid for an analysis of sloppiness using the LIS approach (blue triangles in Figure \ref{fig:relative importance}), when compared to the posterior covariance method (green squares in Figure \ref{fig:relative importance}). For example, the relative eigenvalue size of the tenth posterior covariance eigenvalue is of similar magnitude to that of the third LIS eigenvalue (Figure \ref{fig:relative importance}). Additionally, we previously saw that the eigenvector values became inconsistent when $\lambda_j/\lambda_1 \sim \mathcal{O} (10^{-2})$ using the posterior covariance method (Supplementary Material Section S.5, tenth eigenparameter), so it is not unreasonable to expect inconsistency between samples following the second eigenparameter using the LIS method, when this order of magnitude approximation is similar. Such inconsistencies are expected when eigenvalues are very small because the model-data fit is less sensitive to the corresponding sloppy eigenparameters, and these sloppy eigenparameters are therefore difficult to uniquely identify. Hence, the results of the analysis of model sloppiness used here appears to be consistent across both the LIS and posterior covariance approaches.  

A similar result between the posterior covariance and LIS methods for the stiffest eigenparameters indicates that the prior distribution used for the application of Bayesian inference does not substantially influence on the shape of the posterior distribution. Hence, in our case we concluded that the prior used was weakly informative (as intended) and had little influence on the posterior distribution.

\subsection{Model reduction for the coral calcification model}
The analysis of model sloppiness revealed the parameter combinations which most and least inform model predictions for the coral calcification case study -- a useful tool for identifying potential model simplifications that have little impact on model predictions for the given dataset. {\edit Each analysis of model sloppiness (Section \ref{sec: Sloppiness results}) indicated that the coelenteron-ECM paracellular diffusion (characterised by parameter $k_{pp}$) and the Ca-ATPase pump mechanisms contribute substantially to the stiffest eigenparameter (Figures \ref{fig:MLE original}--\ref{fig:LIS original}), suggesting that these two mechanisms (characterised by ten parameters, see Table \ref{tab: Mechanisms})} have a strong influence on model behaviours. However, the analysis indicates that model predictions are relatively insensitive to the  seawater-coelenteron diffusion mechanism (characterised by parameter $s$), the BAT pump mechanism {\edit (characterised by seven parameters, see Table \ref{tab: Mechanisms})}, and the coelenteron-ECM transcellular diffusion mechanism (characterised by parameter $k_{CO2}$) {\edit because they did not contribute to the stiff eigenparameters (see the first eight parameter combinations in Figure \ref{fig:PCA original})}. Hence, the results suggest that each of these three mechanisms, or a combination of them could be removed. 

However, the seawater-coelenteron diffusion mechanism is an integral part of the model, as it is the primary mechanism by which ocean acidification affects coral calcification rates in the model. Since this model's original purpose in \citet{Galli_2018} was to identify the way that environmental factors affect calcification rates (including but not limited to ocean acidification), this result could suggest that the experimental dataset did not sufficiently capture the effects of ocean acidification on calcification rates. Regardless of this issue, removing the diffusion between the coral polyp and external seawater seems nonsensical as it would yield a model whereby there is no interaction between the local carbon chemistry of seawater and the coral host animal (except indirectly through effects on net photosynthesis which is unlikely; see, e.g.\ \citeauthor{kroeker_2013} \citeyear{kroeker_2013}). 

{\edit It was biologically unreasonable to remove the seawater-coelenteron diffusion mechanism, so the} potential simplified models could include one without the BAT pump, one without the coelenteron-ECM transcellular diffusion, and one without both mechanisms (see Figure \ref{fig:Conceptual model} for a conceptual depiction of these mechanisms). All three of these reduced models were investigated by removing one or both of the selected mechanisms from the ordinary differential equations. The reduced models were recalibrated via Bayesian inference using the same dataset (see Table 1 of \citeauthor{Galli_2018}, \citeyear{Galli_2018}) and likelihood function (Equation \ref{eq:likelihood}), and for the remaining parameters the same independent uniform distributions were used as the prior distribution (Figure \ref{fig:LMLE marginals} and Table S1). Removing both the BAT pump and coelenteron-ECM transcellular diffusion mechanisms from the model reduces the number of parameters from 20 to 12. 

\begin{figure}[H]
    \centering
    \includegraphics[width=0.78\textwidth]{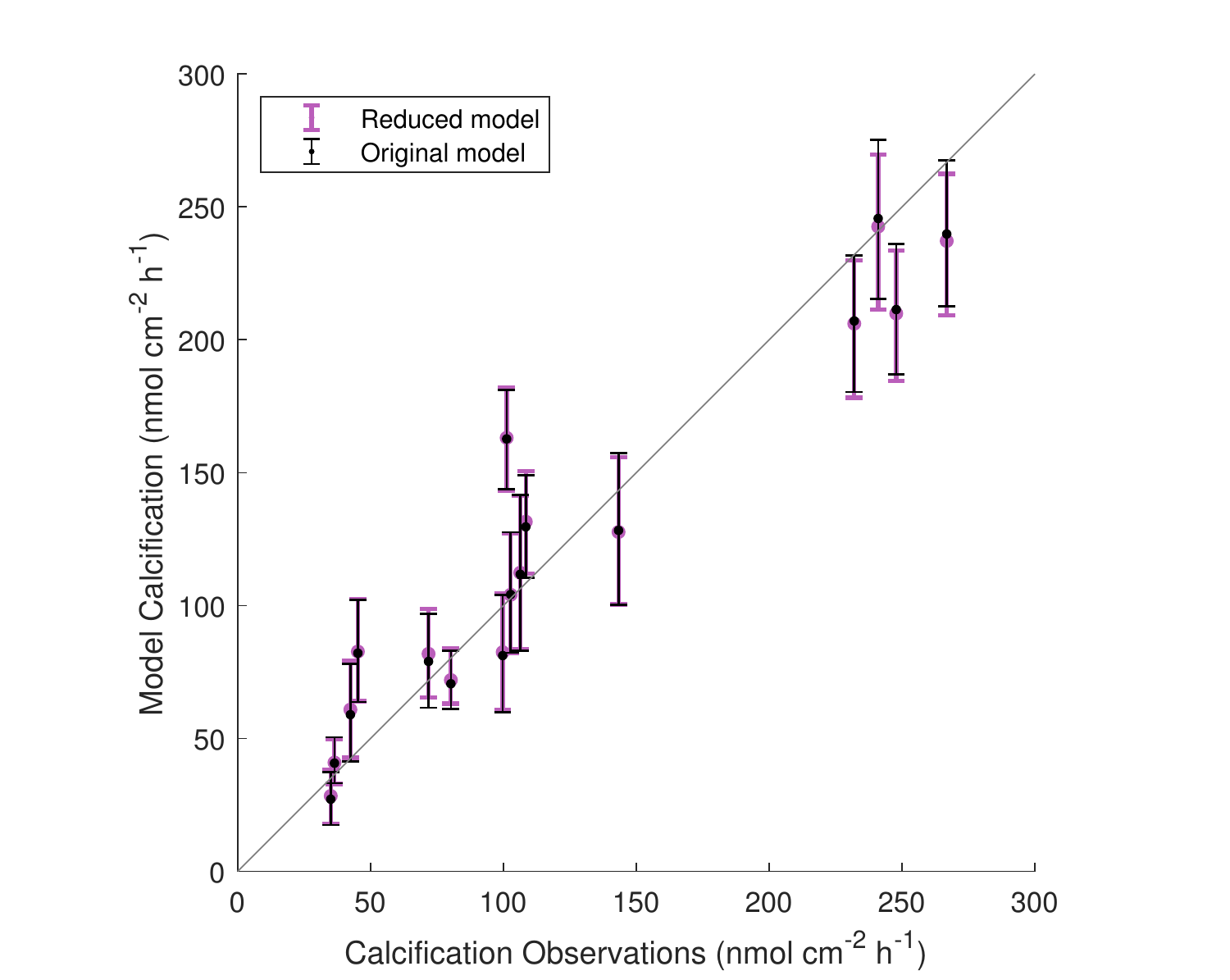}
    \caption{Posterior predictive distribution showing observed calcification compared to modelled calcification for both the original and reduced model. The reduced model is one without both the BAT pump mechanism and coelenteron-ECM transcellular diffusion. The dots indicate the median prediction value, and error bars indicate the 95\% central credible interval for calcification predictions from the respective posterior samples of each model.}
    \label{fig:Goodness of fit comparison}
\end{figure}

{\edit The reduced models suggested by the analysis of model sloppiness yielded very similar predictions of the coral calcification rate data when compared to the original model, despite having up to eight parameters removed. Visually, the goodness-of-fit between the model and data was similar for the original and our proposed reduced model without the BAT pump and coelenteron-ECM transcellular diffusion mechanisms (Figure \ref{fig:Goodness of fit comparison}) -- and a similar result is observed when only one of the two mechanisms were removed (Supplementary Material Section S.7). The estimated model evidence quantitatively suggests that the original model and its three proposed reductions are similarly supported by the data, and the estimated Bayes factors \citep{Kass_1995_BF} do not indicate a strong preference between models (Supplementary Material Section S.8). 

Additionally, removing the two insensitive mechanisms (BAT pump and coelenteron-ECM transcellular diffusion) from the model did not lead to clear differences in the estimated marginal posterior densities for each parameter, or to the analysis of model sloppiness between the original model and the reduced model (Supplementary Material Section S.9). Whilst there is limited sacrifice in predictive ability, there was a significant gain in computation time -- the reduced model required less than half the computation time of the original model for calibration (13.2 hours for the original model and 6.0 hours for the reduced model, using a high-performance workstation with 12 cores).

For comparison, we also removed a mechanism from the model that was represented within the stiffest eigenparameter and therefore very sensitive to the model-data fit. Removing this sensitive mechanism resulted in a much worse model-data fit, both visually (Supplementary Material Section S.7) and quantitatively (Supplementary Material Section S.8). For this case study, these results indicate that analysing the model sloppiness is an appropriate way to inform model reductions for maintaining a good fit between the model and calibration dataset. 
}

\section{Discussion and Conclusion}
\label{sec: Discussion}
In this work, we have proposed and demonstrated a new method for simplifying models based on the analysis of model sloppiness. This analysis can identify the informativeness of parameter combinations on model behaviours by analysing the topology of the surface which describes the fit of the model to data in parameter space \citep{Gloria_2021_sloppy}. {\edit As such, it can identify insensitive model mechanisms whose parameters contribute little to the model's ability to fit the available data, whilst accounting for parameter interdependencies.}
We showed that identifying and removing such insensitive mechanisms can be used for model reduction whilst minimising the loss of predictive capability. 

\subsection{Coral calcification case study}
Our model reduction method, informed by the analysis of model sloppiness, was demonstrated on a case study on a model of coral calcification reaction rates \citep{Galli_2018}. In this case study, the data was not sufficient to provide narrowly constrained estimates for most of the model parameters, but it was also not immediately clear how the model could be simplified. To address this issue, the proposed analysis identified two mechanisms within the model which only weakly informed predictions and were sensible to remove; removing these processes reduced the number of model parameters from 20 to 12. A comparison of the goodness-of-fit for both the original and simplified models indicated that similar predictions were produced by both (Figure \ref{fig:Goodness of fit comparison}). {\edit It should be noted here that the simplified model has, to our best knowledge, not been considered in the literature for modelling coral calcification, so the outcome from analysis of sloppiness yields a new model for practitioners to consider.}

For the present case study, the analysis of model sloppiness may also be useful for better understanding the physiology of the coral polyp. The analysis indicated that the model-data fit was highly sensitive primarily (i.e.\ first eigenparameter) to the Ca-ATPase pump (through parameters $v_{H_c},E0_c,k_{1f_c},k_{2f_c}$ and $k_{1b_c}$) relative to the paracellular diffusion of chemical species (through parameter $k_{pp}$) between the coral polyp compartments that the pump mechanism connects -- the coelenteron and ECM (see Figure \ref{fig:Conceptual model}). This may be expected since it is the balance of the active and passive fluxes between the ECM and coelenteron at steady state that determines the aragonite saturation state of the ECM, which itself directly controls the calcification rate. Secondarily (i.e.\ second eigenparameter), the model-data fit was sensitive to the ATP availability from gross photosynthesis and respiration (through parameters $\alpha$ and $\beta$). In the model this process determines the amount of ATP available to fuel the Ca-ATPase pump responsible for the active flux.  

From our results, we also gain insights from the mechanisms which are removed from the model. Given that the analysis of model sloppiness is closely related to the concept of parameter identifiability \citep{Chis_2016,Browning_2020,Raue_2009_identifiability,Bellman_1970_identifiability}, it can inform the structural and/or practical identifiability of parameters. In the explored coral case study, the removed mechanisms had similar functions to other mechanisms in the model, such that these processes are partially structurally unidentifiable. The BAT pump mechanism is an active transport for bicarbonate anions (HCO$^-_3$), and so the mechanism alters the dissolved inorganic carbon (DIC) levels between the coelenteron and ECM in a similar way to the Ca-ATPase pump. Additionally, the model includes two passive transport mechanisms between the coelenteron and ECM (see Figure \ref{fig:Conceptual model}): a paracellular pathway for all chemical species and a transcellular pathway for carbon dioxide, which have very similar functions. Hence, the analysis of model sloppiness here indicates that certain components of the model are not necessary for a good model-data fit because their functions are partially or fully replicated through other mechanisms. So, when considering the aggregate behaviour of these processes within the coral polyp, including these ``redundant'' physiological processes may not lead to better predictions because of structural and/or practical identifiability issues, and instead may cause poor parameter estimation \citep{Raue_2009_identifiability}. 

Importantly, any physiological conclusions drawn from the results of this analysis are in relation to this dataset alone. The dataset used for this analysis is small, given that the number of data points is of the same order of magnitude as the number of model parameters. {\edit Additionally, any misspecification or inaccurate assumptions in this model are carried through to the reduced model. That being said, all models are unavoidably approximate representations of the real world. } Hence, any model mechanisms excluded through model reduction cannot be considered unimportant for coral reef physiology, as the analysis indicates that model \textit{predictions} of calcification rates for this dataset are negligibly affected by exclusion of these mechanisms. We also acknowledge that these two criticisms are common to data-driven methods which derive simplifications from an originally more complicated model \citep{Crout_2009}.  

The explored coral calcification case study demonstrated the potential for an analysis of model sloppiness to inform model reductions. However, this method could be used generally for strategically proposing simpler models. The method is restricted to parametric models where an appropriate sensitivity matrix can be defined, and is best suited to models where mechanisms can be easily removed (e.g.\ process-based models). However, future work could examine how this model reduction technique performs on various other types of models. For instance, \citet{Gloria_2021_sloppy} describes how an analysis of model sloppiness could be used to identify critical parameter combinations in a stochastic setting, and this idea could potentially be explored for application to model reduction.

\subsection{Sensitivity matrix selection}

In this paper, we compared the results of an analysis of model sloppiness using three different approaches to constructing the sensitivity matrix (Table \ref{tab: Sloppy approaches}). {\edit For the explored case study, all approaches -- the Hessian evaluated at MLE, the posterior covariance method and the LIS method -- agreed on which mechanisms strongly inform the model-data fit (Figures \ref{fig:MLE original}--\ref{fig:LIS original}), so the model reduction informed by each approach was equivalent in this case. However, this conclusion will not always hold for other model-data fitting problems (e.g.,\ see \citeauthor{Gloria_2021_sloppy} \citeyear{Gloria_2021_sloppy}). Just as each approach to capturing the model-data sensitivities differ, so might the model reductions informed by this analysis.} The results from our case study lead to two key general findings regarding selection of a sensitivity matrix. 

Firstly, this case study demonstrated that the non-Bayesian sensitivity matrix may have limited utility when the likelihood surface is rugged. In the coral calcification case study, the likelihood surface had no well-defined peak in parameter space, so single parameter estimates of this peak based on MLE could not reproducibly encapsulate the model-data fit sensitivity. Results based on the local sensitivity of the likelihood surface may not capture important features of the model-data fit in such circumstances, and should be interpreted with caution. For complex models, the global likelihood surface including the full range of parameter values represented by the posterior distribution should instead be considered (e.g.\  using the posterior covariance approach).

Though the Hessian sensitivity matrix may become problematic for some complex models, that does not mean that the MLE Hessian approach cannot be used to gain useful information for model reduction. Rather, a non-Bayesian method of analysing sloppiness is easier to implement and far more computationally efficient. Thus, a non-Bayesian method of analysing sloppiness {\edit -- such as the Hessian matrix evaluated at MLE, or the Levenberg-Marquardt approximation \citep{Marquardt_1963} of the Hessian (used for computationally intensive models) --} could provide a simpler, faster method of suggesting model reductions, in place of the Bayesian counterparts. However, where the likelihood surface is expected to be complicated, sufficient consideration should be given to the choice of optimisation algorithm used to identify the best-fit parameter values. 

Secondly, the case study demonstrated the benefits of using the posterior covariance and LIS methods together. In the explored coral calcification model, both the posterior covariance and LIS results for analysing sloppiness were similar, indicating that the prior distribution was not having a substantial influence on the model-data calibration process (i.e.\ on the posterior distribution). Hence, this case study demonstrates the value of using multiple methods for analysing sloppiness, as together these methods provide richer information than each method by itself \citep{Gloria_2021_sloppy}.

However, because in our case study the prior distribution was weakly informative on the posterior, it is difficult to state what the outcomes of informing model reductions via a LIS analysis would show if informative priors were instead used. Although we leave this exploration for future work, we hypothesise that model reduction informed by LIS could yield reduced models based purely on retaining mechanisms for which the data is highly informative relative to prior beliefs for each mechanism. At the very least, the LIS method is a useful check to identify the influence of the data on the model calibration process within a Bayesian framework.   

{\edit While we only considered three sensitivity matrices, there are various approaches in the literature that could be used within this model reduction framework. Methods such as the Levenberg-Marquardt Hessian, a likelihood-free approximation of LIS, or the active subspace method each capture the model-data fit differently to the methods considered here and may provide advantages for different applications -- such as computationally intensive or stochastic models \citep{Gloria_2021_sloppy}. 

The active subspace method \citep{Constantine_2016_AS} is a dimension reduction approach that constructs a sensitivity matrix based on the gradient of the log-likelihood relative to the prior distribution, capturing informativeness of the data relative to the prior on the model-data fit similarly (but not exactly) as the LIS approach does \citep{Zahm_2022}. \cite{Constantine_2016_AS} define the active/inactive subspace of eigenvectors via eigendecomposition of the sensitivity matrix and identifying the first largest gap between eigenvalues. As noted by \cite{Gloria_2021_sloppy}, the active subspace sensitivity matrix could be used to assess model sloppiness and we suggest it can also be used within our model reduction framework. In addition, \cite{Constantine_2017_AS} proposed activity score metrics to  rank individual model parameters based on the eigenvalues and eigenvectors of the active subspace matrix and can be used for model reduction. We leave explorations of active subspace for model reduction and a comparison with model reductions produced by the sensitivity matrices used in this manuscript for future research.

}

\subsection{Complexity versus parsimony in models}
The processes, phenomena and systems of the world around us are extremely complex; so should the models we create to represent these ideas be equally as complex? The simple answer is that it depends on the purpose of the model, whether that be accurate predictions of collective behaviour \citep{Dietze_2018}, inference for physically meaningful parameters \citep{adams_2017}, or analysis for understanding and/or changing processes within a system (e.g.\ \citeauthor{Galli_2018} \citeyear{Galli_2018}, \citeauthor{Verspagen_2014} \citeyear{Verspagen_2014}).

If the aim of the model is accurate prediction, it is the aggregate behaviour of a system that we aim to recreate through models, and in many cases, including additional underlying processes is not beneficial in modelling the collective behaviours \citep{Transtrum_2014_MBAM,Machta_2013}. Including many potential mechanisms within a model may mean that some processes are being fit to noise in the data, leading to poor predictions in new situations \citep{Cox_2006}. However, a general model may be too simple for accurate prediction and could distort the importance of processes in the model \citep{Lawrie_2007,vanNes_2005}. {\edit Models should minimise the bias-variance tradeoff, balancing the complexity so that the model is simple enough to predict new data well and complex enough to capture features of the data \citep{Geman_1992_biasvariance}.}

What if accurate parameter inference for meaningful parameters is the goal? The more complex a model is, the more difficult parameterisation becomes \citep{vanNes_2005}, and more data is required as a consequence. If unidentifiable parameters are included in the model calibration problem, the values of these parameters become meaningless as they are often interdependent on others \citep{Van_2009}. On the other hand, removing important and existing processes from a model could mean that calibrated parameters lose their physical meaning as they compensate for processes missing from the model and become more like aggregate parameters for modelling the collective model dynamics \citep{Elevitch_2020}.

If there are specific processes within the model that we wish to understand or change within a system, the relevant process needs to be included for analysis \citep{Hannah_2010}. However, even in this case the modeller must still consider whether the parameters retain their physical meaning (due to structural identifiability issues or data limitations) as well as the potential for overfitting caused by the inclusion of the process (e.g.\ \citeauthor{Jakeman_1993}, \citeyear{Jakeman_1993}). Additionally, there are other issues that come with complex models, such as difficulty implementing, reproducing, interpreting, validating and communicating the models, as well as computation times and difficulty or costs associated with updating or replacing these models \citep{vanNes_2005}.

So where is the line between too simple and too complex? There are many arguments both for and against complexity (e.g.\ \citeauthor{Anderson_1972}, \citeyear{Anderson_1972}; \citeauthor{Hong_2017}, \citeyear{Hong_2017}; \citeauthor{Hunt_2007}, \citeyear{Hunt_2007}; \citeauthor{Logan_1994}, \citeyear{Logan_1994}; \citeauthor{Wigner_1990}, \citeyear{Wigner_1990}). {\edit The desired complexity of a model should be based on both the model’s purpose and statistical assessments of quality \citep{Saltelli_2019}}. Our goal in the present work is to highlight that, in some circumstances, model reduction could benefit the predictions and parameter estimates of a model. Our model reduction framework provides a principled and intuitive approach for model simplifications to address this goal. 

\subsection*{Software and data availability}
The code used for this analysis was implemented in MATLAB (R2021b), and is freely available for download on Figshare at \url{https://doi.org/10.6084/m9.figshare.19529626.v2}. This code (14.8MB) contains 37 functions as MATLAB code files, which runs the analyses for the coral calcification case study and produces the corresponding plots described within this manuscript. In addition, 7 MATLAB data files are included, which contain the dataset used for analyses (available to access through \cite{Galli_2018}), and generated by the analysis (including SMC samples for each model, and for independent runs, local MLE samples, and calculated sensitivity matrices).   

\subsection*{Declaration of competing interest}
The authors declare that they have no known competing financial interests or personal relationships that could have appeared to influence the work reported in this paper.

\subsection*{Acknowledgements}

The authors thank Catherine Kim for helpful discussions during development of the manuscript. Computational resources were provided by the eResearch Office, Queensland University of Technology.

\textbf{Funding}: SAV is supported by a Queensland University of Technology Centre for Data Science Scholarship. CD is supported by an Australian Research Council Future Fellowship (FT210100260). MPA and SAV acknowledge funding support from an Australian Research Council Discovery Early Career Researcher Award (DE200100683). GMM-B acknowledges funding support for an Australian Research Council Linkage Grant (LP160100496) and a Research Stimulus (RS) Postdoctoral Fellowship from the University of Queensland.

\textbf{Author contributions}: SAV performed the numerical experiments and wrote the first draft of the manuscript; SAV, CD and MPA contributed to the design of the research; all authors contributed to the coding, analysis of results, review and editing of the manuscript.

\newpage
\bibliographystyle{chicagoa}
\bibliography{Refs}

\end{document}

\begin{center}
    \appendix
\section*{Supplementary Material}
\end{center}

\renewcommand{\baselinestretch}{1.15}
   \begin{center}
      \Large\textbf{Strategic model reduction by analysing model sloppiness: a case study in coral calcification}\\
      \large{Sarah A. Vollert$^{a,b,c,*}$, Christopher Drovandi$^{a,b,c}$, Gloria M. Monsalve-Bravo$^{d,e,f}$ \& Matthew P. Adams$^{a,b,c,d}$} \\

      \footnotesize
$^a$Centre for Data Science, Queensland University of Technology, Brisbane, QLD 4001, Australia \\

$^b$ARC Centre of Excellence for Mathematical and Statistical Frontiers, Queensland University of Technology, Brisbane, QLD 4001, Australia \\

$^c$School of Mathematical Sciences, Queensland University of Technology, Brisbane, QLD 4001, Australia\\

$^d$School of Chemical Engineering, The University of Queensland, St Lucia, QLD 4072, Australia\\

$^e$School of Earth and Environmental Sciences, The University of Queensland, St Lucia, QLD 4072, Australia\\

$^f$Centre for Biodiversity and Conservation Science, The University of Queensland, St Lucia, QLD 4072, Australia\\

$^*$Corresponding author. E-mail: sarah.vollert@connect.qut.edu.au

   \end{center}

\normalsize
\setlength{\parskip}{1em}

\renewcommand{\thesubsection}{S.\arabic{subsection}}

\setcounter{table}{0}
\renewcommand{\thetable}{S\arabic{table}}

\setcounter{figure}{0}
\renewcommand{\thefigure}{S\arabic{figure}}

\setcounter{equation}{0}
\renewcommand{\theequation}{S\arabic{equation}}

\subsection{Further details of the coral calcification model}
\label{SM: Coral model}
The deterministic model proposed by \cite{Galli_2018} is used to predict the rate at which calcification occurs based on the steady-state solution of a system of nonlinear ODEs. There are two modelled compartments within the coral polyp, the coelenteron, which is a fluid compartment considered to be the stomach of the polyp, and the extracellular calcifying medium (ECM), the compartment where calcification reactions occur within the coral. Additionally, the model also includes the surrounding seawater, as seawater exchanges chemical species with the coelenteron of the coral polyp. 

As a result of the reactions occurring within the coral polyp, as well as exchanges of species between the compartments, the ODE model is comprised of flux terms that adjust the chemical species concentrations. Calcification is one such reaction, where calcium ions (Ca$^{2+}$) and carbonate ions (CO$_3^{2-}$) precipitate into calcium carbonate (CaCO$_3$) in the ECM. The process of calcification is mathematically described as a flux term within the system of ODEs. Calcification rates depend solely on the concentration of the relevant chemical species (Ca$^{2+}$ and CO$_3^{2-}$), so prediction of calcification rates requires knowledge of the concentrations of the chemical species in each compartment.

The model also assumes the presence of the enzyme carbonic-anhydrase within the coral host, so that it catalyses the equilibration of the carbonate system \citep{Bertucci_2013}. Therefore, it is assumed that the carbonate system is at separate equilibria within both the coelenteron and ECM. This assumption allows the entire carbonate system to be described uniquely by knowledge of any two chemical species of the carbonate system. Hence, the carbonate system is modelled using only dissolved inorganic carbon (DIC) and total alkalinity (TA), and equations describing carbonate species equilibrium relationships are used to calculate the concentration of all other carbonate system species (H$^+$, OH$^{-}$, CO$_2$, HCO$^{-}_3$ and CO$_3^{2-}$). 

The various flux terms, which either represent the flow of chemical species between compartments or chemical reactions, are derived from current understanding of coral physiology. Justification and description of the mathematical forms of the ODEs and these flux terms are provided in \citet{Galli_2018}; here we summarise the purpose of these fluxes and the parameters requiring estimation that characterise them. There are four types of processes that are assumed to control the carbonate system and calcium ion concentrations within the coral host:
\begin{enumerate}
    \item \textbf{Gross photosynthesis and respiration reactions} -- These are the processes of carbon exchange due to the zooxanthellae algae which are in symbiosis with the coral host. Photosynthesis and respiration reactions remove and supply DIC from the coral host respectively, and both reactions produce ATP energy. In the model of \cite{Galli_2018}, the chemical fluxes due directly to photosynthesis and respiration are treated as model inputs, so do not have corresponding free parameters that require estimation. 
    \item \textbf{Passive transport} -- There are three routes for chemical species to passively diffuse between the coral compartments within the polyp and the external seawater. These three passive transport mechanisms are:
    \begin{enumerate}
        \item Coelenteron-ECM transcellular diffusion: a transcellular pathway for carbon dioxide between the coelenteron and ECM (characterised by parameter $k_{CO2}$),
        \item Coelenteron-ECM paracellular diffusion: a paracellular pathway for all chemical species between the coelenteron and ECM (characterised by parameter $k_{pp}$), and
        \item Seawater-coelenteron diffusion: a passive exchange mechanism between the coelenteron and external seawater (characterised by parameter $s$).
    \end{enumerate}
    \item \textbf{Active transport (Ca-ATPase and BAT pumps)} -- The model assumes there are two active pump mechanisms between the coelenteron and ECM compartments of the coral polyp, which each increase the aragonite saturation state of the ECM. 
    \begin{enumerate}
        \item The Ca-ATPase pump increases the calcium ion concentration in the ECM in exchange for protons being transported to the coelenteron. This pump mechanism is assumed to be driven by ATP energy sourced from photosynthesis and respiration fluxes, so is characterised by 10 parameters requiring estimation ($\alpha, \beta$ for the ATP energy budget, and the 8 Ca-ATPase parameters indicated in Table \ref{tab:Parameter Summary}).
        \item The BAT pump controls the movement of bicarbonate anions (HCO$_3^-$) between the coelenteron and ECM, and is characterised by 7 parameters requiring estimation (see Table \ref{tab:Parameter Summary}).
    \end{enumerate}
    \item \textbf{Calcification reactions} -- The key process rate which is being predicted. The rate of this reaction depends on concentration of calcium and carbonate ions in the ECM; the parameters associated with this reaction do not require estimation in the model of \citet{Galli_2018}.
\end{enumerate}

The coral compartments and flux terms associated with these four processes are visualised in Figure 3 of the manuscript. These underlying components of the model are combined together into a system of ordinary differential equations (see Equations (23--28) in \citet{Galli_2018}). This system models the flow of calcium ions, DIC and TA throughout the coral polyp. To keep the carbonate species in equilibrium at each numerical timestep of the ODE, the pH and carbonate species concentrations (H$^+$, OH$^{-}$, CO$_2$, HCO$^{-}_3$ and CO$_3^{2-}$) are recalculated based on the current DIC, TA, temperature and salinity using the MATLAB version of CO2SYS \citep{Heuven_2011}. This CO2SYS algorithm is a commonly used software package for calculating carbonate species equilibrium concentrations; full details of this algorithm are provided in \cite{Orr_2015}. The ODE model in tandem with the CO2SYS algorithm in the coelenteron and ECM compartments are together solved at steady state, so that the concentrations of species and fluxes have stabilised, to gain an estimate for the coral polyp's calcification rate. Such stabilisation is expected to occur over timescales of an hour or less \citep{tambutte_1996,al_horani_2003}.

\newpage
\subsection{Coral calcification model parameters}
\label{SM: parameter table}

Parameters of the coral calcification model that are estimated in this work via MLE and Bayesian inference are summarised in Table \ref{tab:Parameter Summary}, with assumed values indicated where possible. 

\begin{table}[H]
\scriptsize
    \centering
    \begin{tabular}{m{0.1\textwidth}|m{0.09\textwidth}|m{0.22\textwidth}|m{0.1\textwidth}|m{0.13\textwidth}|m{0.2\textwidth}}
         \textbf{Mechanism} & \textbf{Parameter} & \textbf{Description} & \textbf{Value Range} & \textbf{Units} & \textbf{Source/Justification of Range} \\ \hline
         Coelenteron-ECM transcellular diffusion  & $k_{CO2}$ & CO$_2$ permeability constant & 0-0.1 & cm~s$^{-1}$ & Strictly positive, maximum velocity should not exceed passive diffusion of seawater \citep{Zeebe_2011} \\ \hline
         Coelenteron-ECM paracellular diffusion & $k_{pp}$ & Paracellular pathway permeability & 0-0.1& cm~s$^{-1}$ & Strictly positive, maximum velocity should not exceed passive diffusion of seawater \citep{Yuan_1974,Zeebe_2011}\\ \hline
         Seawater-coelenteron diffusion & $s$ & Diffusion coefficient & 0-0.1 & cm~s$^{-1}$ & Same as $k_{pp}$\\ \hline
         ATP energy budget (for Ca-ATPase mechanism) & $\alpha$ & Fraction of Pg allocated to calcification & 0-1 & - & Fraction must be between 0 and 1 \\
         & $\beta$ & Fraction of R allocated to calcification & 0-1 & - & Fraction must be between 0 and 1 \\ \hline
         Ca-ATPase mechanism & $v_{H_c}$ & Proportionality constant (Ca-ATPase) & 0-250 & cm~s$^{-1}$ & Strictly positive \\ 
         & $E0_c$ & Ca-ATPase concentration & 0-1.2$\times10^{7}$ & $\mu$mol~cm$^{-2}$ & Strictly positive\\
         & $k_{1f_c}$ & Ca-ATPase rate constant & 0-1.4$\times10^{-4}$ & cm$^4$~s~$\mu$mol$^{-2}$ & Strictly positive\\
         & $k_{2f_c}$ & Ca-ATPase rate constant & 0-0.5 & cm$^2$~$\mu$mol$^{-1}$ & Strictly positive\\
         & $k_{3f_c}$ & Ca-ATPase rate constant & 0-800 & s$^{-1}$ & Strictly positive \\
         & $k_{1b_c}$ & Ca-ATPase rate constant & 0-8 & s$^{-1}$ & Strictly positive\\
         & $k_{2b_c}$ & Ca-ATPase rate constant & 0-500 & s$^{-1}$ & Strictly positive\\
         & $k_{3b_c}$ & Ca-ATPase rate constant & 0-1.0$\times10^{-7}$ & cm$^4$~s~$\mu$mol$^{-2}$ & Strictly positive\\ \hline
         BAT mechanism & $E0_b$ & BAT concentration & 0-1500 & $\mu$mol~cm$^{-2}$ & Strictly positive\\
         & $k_{1f_b}$ & BAT rate constant & 0-5.0$\times10^{-5}$ & cm$^3$~s$^{-1}$~$\mu$mol$^{-1}$ & Strictly positive\\
         & $k_{2f_b}$ & BAT rate constant & 0-0.01 & s$^{-1}$ & Strictly positive\\
         & $k_{3f_b}$ & BAT rate constant & 0-0.01 & s$^{-1}$ & Strictly positive\\
         & $k_{1b_b}$ & BAT rate constant & 0-2.0$\times10^{-4}$ & s$^{-1}$ & Strictly positive\\
         & $k_{2b_b}$ & BAT rate constant & 0-1.0$\times10^{-3}$ & s$^{-1}$ & Strictly positive\\
         & $k_{3b_b}$ & BAT rate constant & 0-3.5$\times10^{-8}$ & cm$^3$~$\mu$mol$^{-1}$~s$^{-1}$ & Strictly positive\\ 
    \end{tabular}
    \caption{Calibrated model parameters as defined by \cite{Galli_2018}. Value ranges stated for estimated parameters represent the bounds for the uniform prior distributions assumed in this work. }
    \label{tab:Parameter Summary}
\end{table}

\newpage
\subsection{Goodness-of-fit of local maxima}
\label{SM: GOF local maxima}

\begin{figure}[H]
    \centering
    \includegraphics[width=\textwidth]{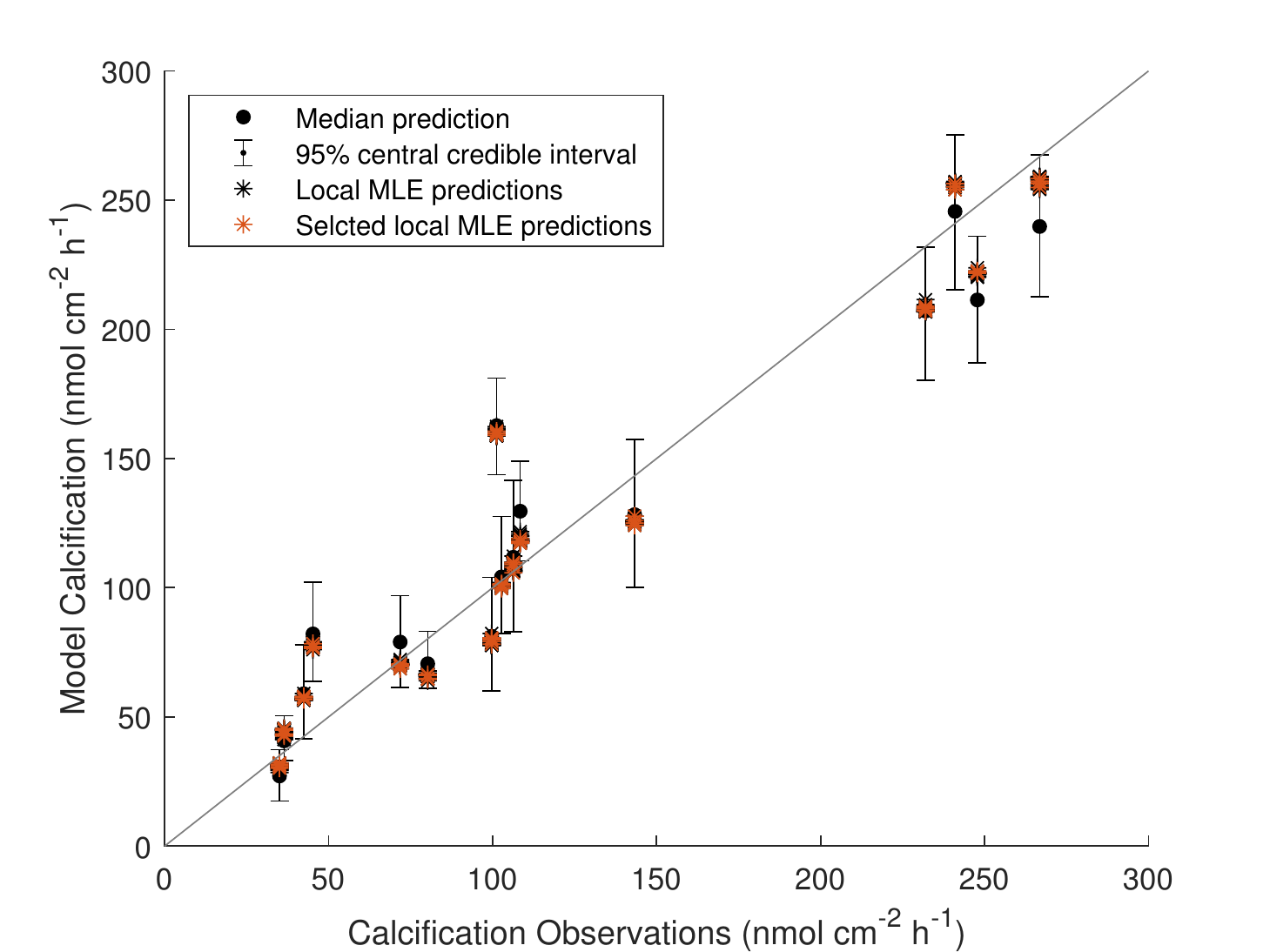}
    \caption{Posterior predictive distribution showing the goodness-of-fit for observed calcification versus model-predicted calcification, from Bayesian inference (black dots and error bars). The local likelihood maxima identified using MLE are shown for 25 parameter estimates (black asterisks) and the five estimates with the highest likelihood for the analysis of model sloppiness are also shown (orange asterisks). Each of the local maxima produce similarly good fits to the data despite having distinctly different parameter values (see corresponding orange and black vertical lines in Figure 4 of the manuscript).}
    \label{fig:LMLE posterior predictive}
\end{figure}

In this paper we found that the likelihood surface for the fit of the coral calcification model to data did not contain a well-defined peak. As such, many local maxima with similarly high likelihoods could be found by initiating a gradient-based search function at different locations in parameter space. These local maxima have distinct parameter values, yet each reflects a similarly good model-data fit. The goodness-of-fit for each local maxima is shown in Figure \ref{fig:LMLE posterior predictive}.

\newpage
\subsection{Further details for calibration of the coral calcification model}

\label{SM: Calibration details}
\subsubsection{Prior distribution}
\label{sec:priors}
When Bayesian inference was applied to this problem, a prior distribution was specified for each of the unknown parameters, 
\begin{align}
    \bm{\theta} &= (k_{CO2},\, k_{pp},\, s,\, \alpha,\, \beta,\, v_{H_c},\, E0_c,\, k_{1f_c},\, k_{2f_c},\, k_{3f_c},\, k_{1b_c}, \nonumber \\
    &\qquad \, k_{2b_c}, \, k_{3b_c}, \, E0_b, \, k_{1f_b},\, k_{2f_b},\, k_{3f_b},\, k_{1b_b},\, k_{2b_b},\, k_{3b_b},\, \sigma),
    \label{Eq: theta}
\end{align}
where $\sigma$ (units of $\mu$mol~cm$^{-2}$~h$^{-1}$) is the standard deviation characterising measurement noise, that is also estimated via Bayesian inference. 

Some of the parameters had clearly definable limits. For example, $\alpha$ and $\beta$ defined the fractions of ATP from gross photosynthesis and respiration utilised by the Ca-ATPase pump, and therefore must possess values between 0 and 1. We could also sensibly surmise that the speed of passive chemical species movement $k_{CO2}, k_{pp}$ and $s$ has a maximum velocity equivalent to passive diffusion within seawater (Table \ref{tab:Parameter Summary}). However, most of the parameters for this model were only known to be strictly positive, yet had no other obvious choice for prior distributions. As there was little information available on the parameters, uniform priors were used for each parameter. There was also no prior knowledge of covariance between parameters, so the joint prior $\pi(\bm{\theta})$ was chosen to be the product of independent uniform priors for each parameter,
\begin{equation}
    \pi(\bm{\theta})=p(k_{CO2})\times \cdots\times p(\sigma), 
\end{equation}
where $p(k_{CO2})$, ..., $p(\sigma)$ are the independent uniform prior distributions for the model parameters $k_{CO2}$, ..., $\sigma$, listed in Equation (\ref{Eq: theta}). The upper and lower bounds of the uniform priors for each parameter were chosen to match the range of values specified in Table \ref{tab:Parameter Summary}. Additionally, $\sigma$ was assumed to have a uniform prior bounded between 0 and 50~$\mu$mol~cm$^{-2}$~h$^{-1}$. 

During preliminary simulations, some steady state solutions of the ODE model did not converge when calculating the carbonate species equilibrium using the CO2SYS algorithm. To deal with this issue, parameter values that caused divergence of CO2SYS were excluded from the prior distribution. This modified the joint prior to 
\begin{equation}
    \pi(\bm{\theta}) \propto p(k_{CO2})\times \cdots\times p(\sigma) \times \textbf{1}_c(\bm{\theta}),
\end{equation}
where the indicator function $\textbf{1}_c$ was defined here to output one if the CO2SYS converges and zero otherwise. This rejection procedure did not substantially alter the shape of the marginal prior distributions (shaded grey regions in Figure 4 of the manuscript). In addition, we found in practice that rejections due to divergence of the CO2SYS algorithm reduced as the SMC algorithm we used for Bayesian inference moved towards areas of high posterior density. Hence, for this coral calcification model, areas of low posterior density are more likely to involve parameter combinations that yield highly unreasonable predictions that subsequently cause divergence of CO2SYS. However, we also cannot rule out the possibility that this issue was due to the numerical procedures we used. Whether this divergence issue was a result of areas of low posterior density or the numerical procedures, the resulting prior distributions which appear uniform, and the observation that the rate of rejection due to divergence reduced as areas of high posterior density were visited more often, suggested that the approach of rejecting these parameter samples was reasonable. 

\subsubsection{Posterior simulation}
\label{sec:posterior}
The SMC algorithm used to estimate the posterior distribution was adapted from \cite{drovandi_2011}. For this application, an SMC algorithm based on a likelihood annealing approach was used. 
This approach allowed the ensemble of particles to steadily converge onto the posterior distribution. 

Within the algorithm, a transform was placed on each of the parameters, such that any real number could be proposed and converted to a proposal for $\bm{\theta}$ within the prior bounds. This transform was applied for each parameter, where the transforms for the $j$th parameter $\theta_j \rightarrow \tilde{\theta}_j$ and the equivalent inverse form $\tilde{\theta}_j \rightarrow \theta_j$ were defined as
\begin{equation}
    \tilde{\theta}_j = \log \left(\frac{\theta_j-l_j}{u_j - \theta_j} \right), \qquad \theta_j = \frac{l_j+u_j e^{\tilde{\theta}_j}}{1+e^{\tilde{\theta}_j}},
\end{equation}
where $l_j$ and $u_j$ are the lower and upper bounds for parameter $\theta_j$, respectively. This transform then allowed any real number proposed as $\tilde{\theta}_j$ to be converted to a $\theta_j$ value within the uniform prior bounds, such that if $- \infty < \tilde{\theta_j} < \infty$, then $l_j \leq \theta_j \leq u_j$. 

The final result of the SMC algorithm was then visually examined to ensure that the results were reproducible over multiple independent runs of the algorithm, such that the posterior distributions were similar across independent sets of samples. For this application, the SMC algorithm was run independently five times to produce five independent sets of posterior samples. The estimated marginal distributions and posterior predictive distributions were visually very similar for each independent set of 5000 samples. Hence, the results were found to be highly reproducible for 5000 posterior samples. Each of these five independent samples were then used to ensure reproducibility of the results of an analysis of model sloppiness, using a posterior covariance or LIS sensitivity matrix.

\newpage
\subsection{Analysis of model sloppiness using the posterior covariance method}
\label{SM: PC reproducibility}

\begin{figure}[H]
    \centering
    \includegraphics[width=12cm,trim={0cm 0.2cm 0cm 0cm},clip]{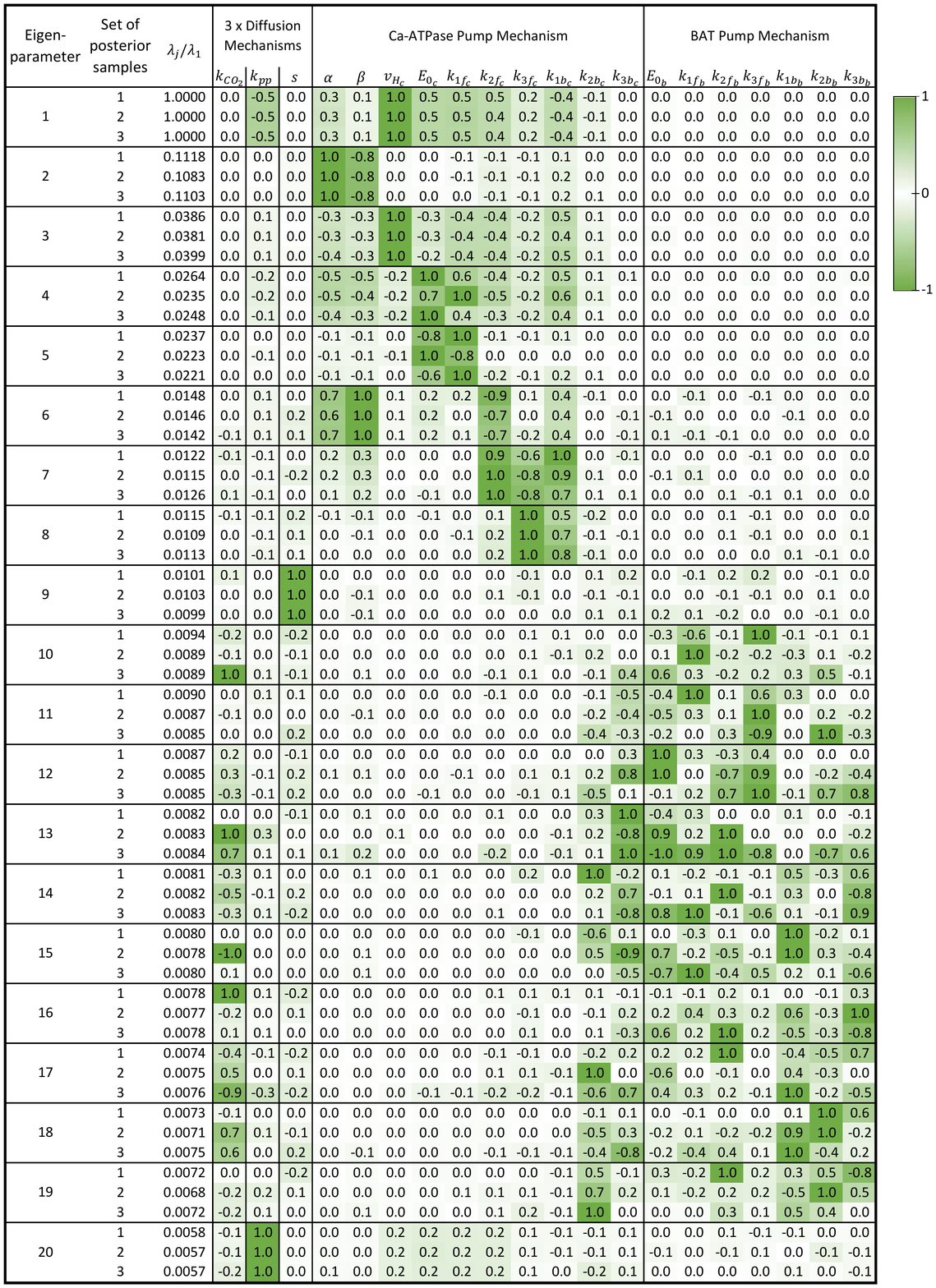}
    \caption{Eigenvector element values for eigenparameters identified using a posterior covariance analysis of sloppiness. These eigenparameters are ordered from highest relative importance to lowest according to the size of the eigenvalues. For each eigenparameter, the results of three independent sets of posterior samples with 5000 particles each are compared to indicate the consistency of results. For each eigenparameter, the values were normalised by the leading eigenvector value, such that they are rescaled to be between -1 and 1 inclusive. Here the colour darkens as the absolute values of the eigenvector values increase from 0 to 1, such that the model-data fit is more sensitive to darker eigenvector values. Additionally, the relative size of each eigenvalue when compared to the leading eigenvalue for each sample has been included in the column $\lambda_j/\lambda_1$.}
    \label{fig: PCA reproducibility}
\end{figure}

Multiple independent results of the analysis of model sloppiness via a posterior covariance sensitivity matrix were produced to analyse the consistency of results. Five sets of posterior samples were independently produced using the SMC algorithm and each of these were used to analyse the sloppiness of the coral calcification model. Figure \ref{fig: PCA reproducibility} shows the eigenparameters for three of the five sets of independent samples, revealing that the results are reproducible for the first nine eigenparameters (the results for the two posterior samples not shown here were similar). After this point, the independent analyses show different model-data fits; however, this can be expected because these parameter combinations commonly become less constrained in a sloppy model.

\newpage
\subsection{Analysis of model sloppiness using the LIS method}
\label{SM: LIS}
\begin{figure}[H]
    \centering
    \includegraphics[width=\textwidth,trim={0cm 3.7cm 3.8cm 0cm},clip]{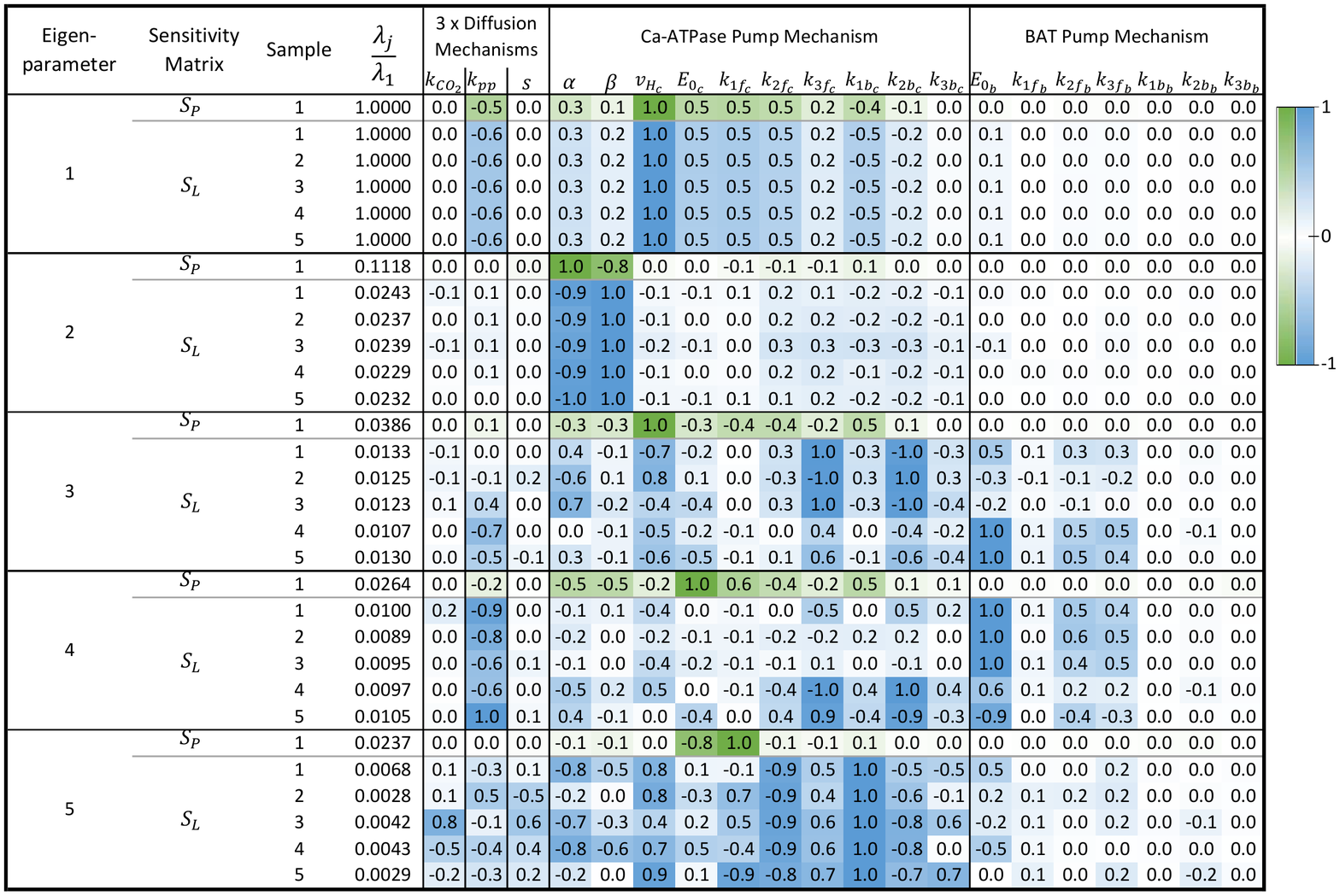}
    \caption{Comparison of eigenvector values of an analysis of model sloppiness using a LIS ($S_L$) and posterior covariance ($S_P$) sensitivity matrix (blue and green respectively). For each LIS matrix eigenparameter, the results of five independent sets of posterior samples with 5000 particles each are compared to indicate the consistency of results. Five eigenvector values are also shown for a posterior covariance method of analysing model sloppiness (green) from a single sample of 5000 particles, since the eigenvector values from this analysis were fairly consistent across independent samples (Figure \ref{fig: PCA reproducibility}). See the caption of Figure \ref{fig: PCA reproducibility} for further interpretation of this figure.}
    \label{fig:LIS reproducibility}
\end{figure}

Here, we compare multiple independent analyses of model sloppiness using a LIS sensitivity matrix to show the consistency of results. Five sets of posterior samples were  independently produced using the SMC algorithm and each of these were used to analyse the sloppiness of the coral calcification model. Figure \ref{fig:LIS reproducibility} shows the eigenparameters for each of these five independent sets of samples, revealing that the results are somewhat reproducible for the two stiffest eigenparameters, but not any others. This figure also shows the results of a posterior covariance method approach for comparison, as this reveals that the two approaches have a similar model-data fit for the two stiffest eigenparameters. 

\newpage
\subsection{Goodness-of-fit comparisons for other reduced models}
\label{SM: GOF all models}

\begin{figure}[H]
    \centering
    \includegraphics{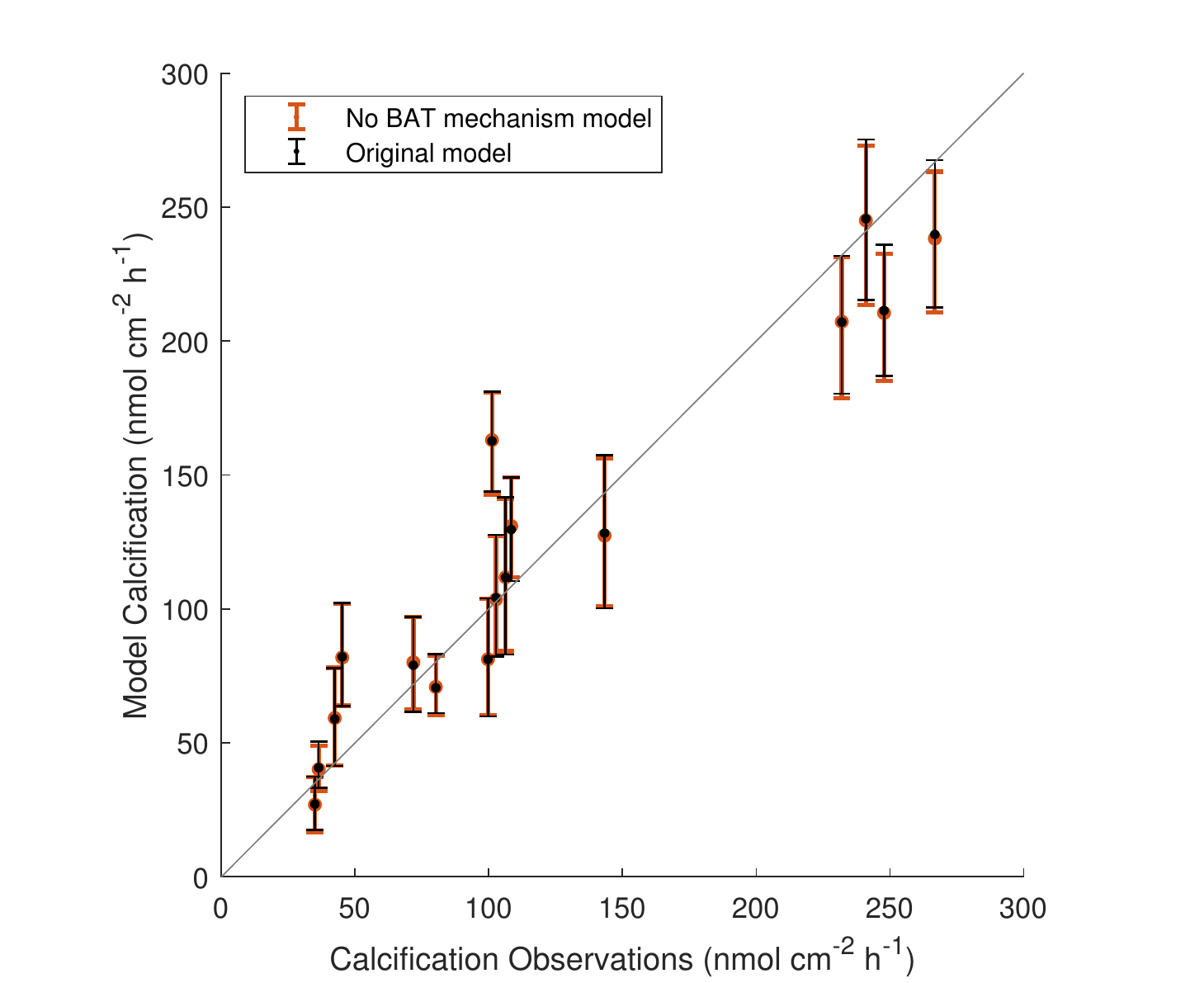}
    \caption{Posterior predictive distribution showing observed calcification compared to modelled calcification for both the original and reduced model. Here the reduced model is one without the BAT pump mechanism. Here, the dots indicate the median prediction value, and error bars indicate the 95\% central credible interval for calcification predictions from the respective posterior samples of each model.}
    \label{fig: No BAT GOF}
\end{figure}

\begin{figure}[H]
    \centering
    \includegraphics{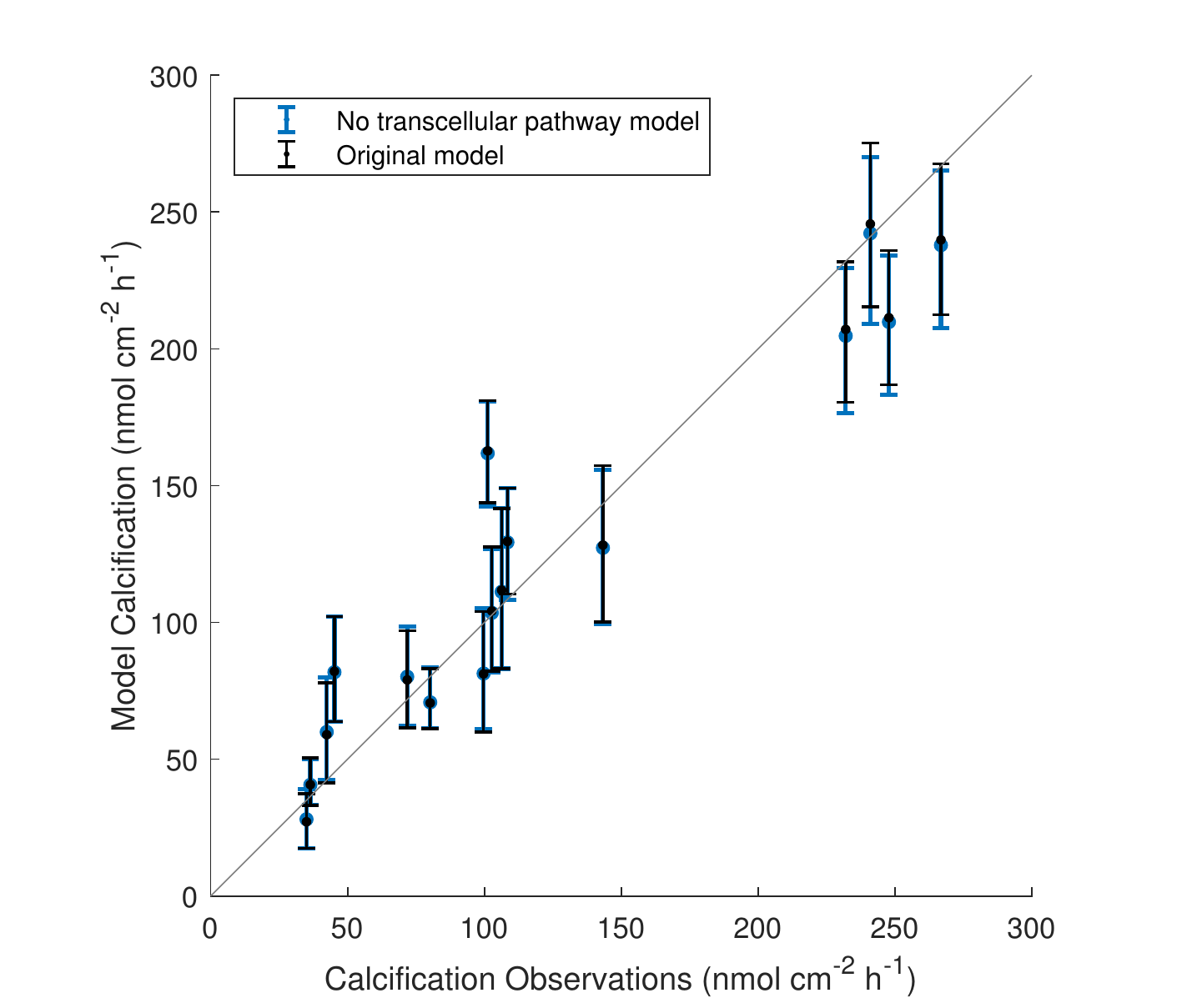}
    \caption{Posterior predictive distribution showing observed calcification compared to modelled calcification for both the original and reduced model. Here the reduced model is one without the coelenteron-ECM transcellular diffusion for CO$_2$. Here, the dots indicate the median prediction value, and error bars indicate the 95\% central credible interval for calcification predictions from the respective posterior samples of each model.}    \label{fig: No trans GOF}
\end{figure}

Firstly, two other model reductions informed by the analysis of model sloppiness were tested: one where only the BAT pump mechanism was removed and one where only the coelenteron-ECM transcellular diffusion was removed. The goodness-of-fit of the model without BAT pump mechanism and the model without the coelenteron-ECM transcellular diffusion are shown in Figures 
\ref{fig: No BAT GOF} and \ref{fig: No trans GOF} respectively.

\begin{figure}[H]
    \centering
    \includegraphics{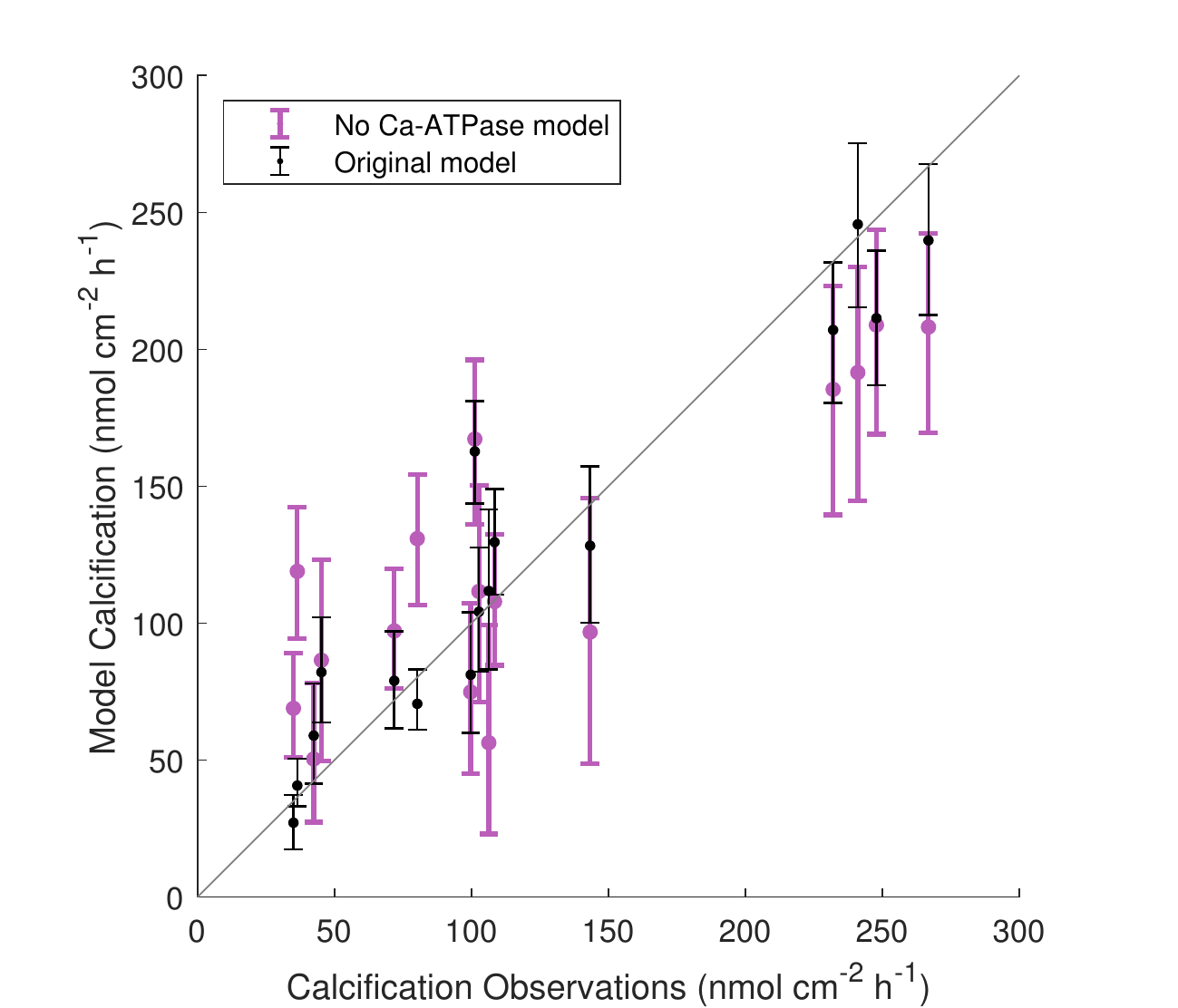}
    \caption{Posterior predictive distribution showing observed calcification compared to modelled calcification for both the original and reduced model. Here the reduced model is one without the Ca-ATPase pump mechanism. Here, the dots indicate the median prediction value, and error bars indicate the 95\% central credible interval for calcification predictions from the respective posterior samples of each model.}    \label{fig: No Ca-ATPase GOF}
\end{figure}

{\edit Secondly, we investigated models where mechanisms necessary for a good model-data fit were removed. The analysis of model sloppiness identified that the coelenteron-ECM paracellular diffusion and Ca-ATPase pump mechanisms contributed substantially to the stiffest eigenparameter (Figure 7 of the manuscript). Hence, we sought to analyse models with these mechanisms excluded as a test case for what happens if the sloppy analysis recommendations for which mechanisms should be removed are not followed. Removing the coelenteron-ECM paracellular diffusion results in the exchange of chemical species between the coelenteron and ECM to being controlled entirely though the exchange of carbon dioxide. Consequently, calibrating this model resulted in numerical issues, and so we did not further consider the model without coelenteron-ECM paracellular diffusion. Analysing the remaining unrecommended model reduction -- removal of the Ca-ATPase pump mechanism -- resulted in a visibly poorer goodness-of-fit in comparison to the original model  (Figure \ref{fig: No Ca-ATPase GOF}). }

\subsection{Bayes' factors for investigated models}
{\edit For each of the models considered in this case study, the Bayesian model evidence was estimated for the sample and used to produce approximate Bayes' factors \citep{Kass_1995_BF} for comparing models to the original model. Values close to one identify that the model evidence between the compared models is similar, whereas a value of 10 or more might suggest strong evidence to prefer the original model. The estimated Bayes' factors quantitatively suggest that each of the model reductions suggested by the analysis of model sloppiness are comparable to the original model (Table \ref{tab:Bayes factors}). However, when looking at a model reduction which was not recommended by the analysis of model sloppiness there is strong evidence to prefer the original model for this dataset.} 

\begin{table}[H]
    \centering \edit
    \begin{tabular}{p{0.60\textwidth}|p{0.33\textwidth}}
       \textbf{Model mechanism(s) removed} & \textbf{Estimated Bayes' factor} \\ \hline
        Coelenteron-ECM transcellular diffusion & $1.12$ \\
        BAT pump & $1.26$ \\
        Coelenteron-ECM transcellular diffusion and BAT pump & $1.25$ \\
        Ca-ATPase pump & $7.30 \times 10^6$ \\
    \end{tabular}
    \caption{The estimated Bayes' factors for each reduced model compared to the original model. Note that the analysis of sloppiness suggests the first three model reductions, however, it suggests that the Ca-ATPase pump is important for a good model-data fit.}
    \label{tab:Bayes factors}
\end{table}

\newpage
\subsection{Reduced model compared to original model}
\label{SM: marginals both models}

\begin{figure}[H]
    \centering
    \includegraphics[width=\textwidth]{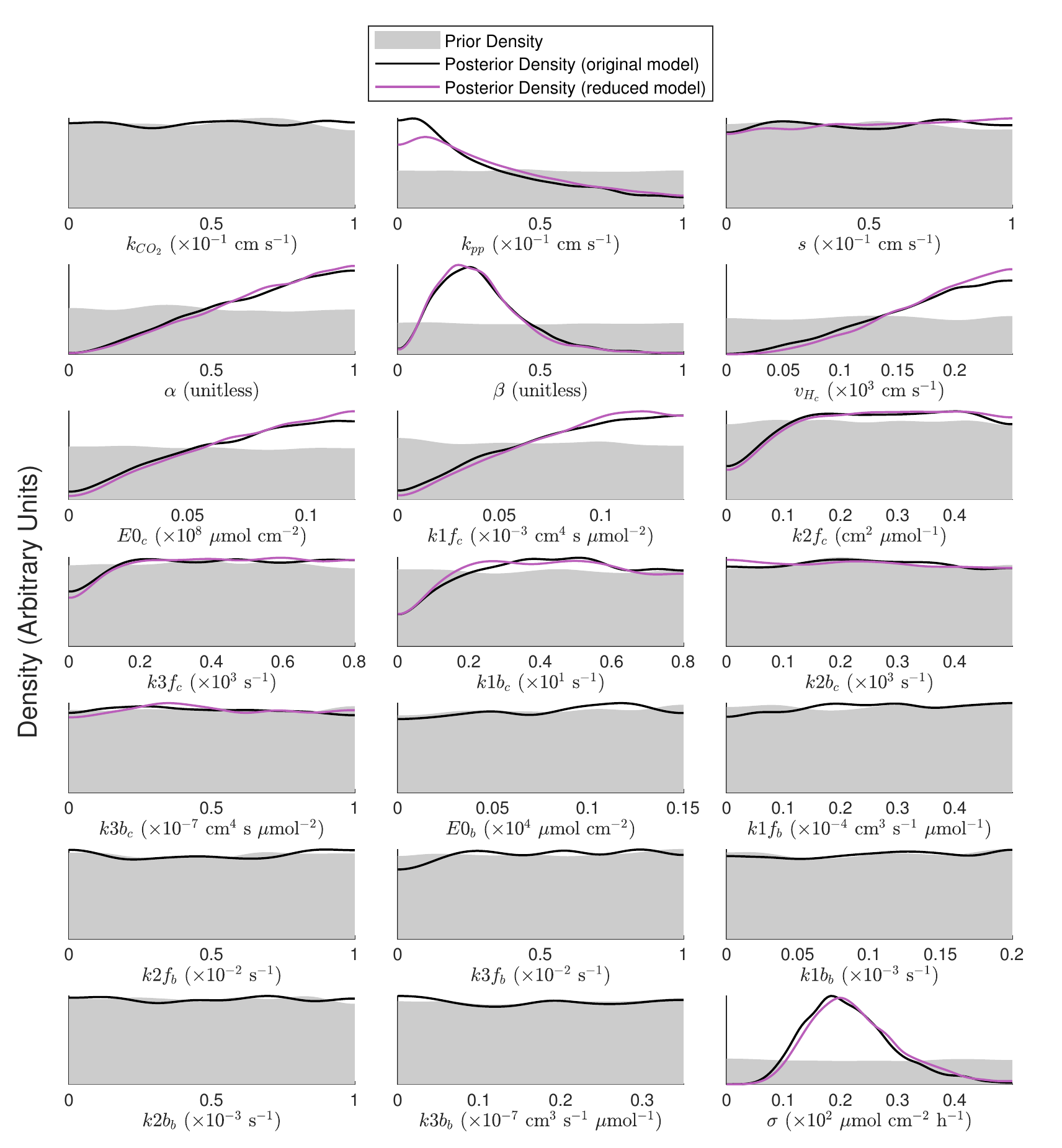}
    \caption{Estimated posterior marginal densities for the model parameters from 5000 posterior samples. This figure shows the uncertainty of model parameters for both the original and reduced model. Parameters $k_{CO_2}, E0_b, k_{1f_b}, k_{2f_b}, k_{3f_b}, k_{1b_b}, k_{2b_b},$ and $k_{3b_b}$ are present in the original model but not in the reduced model, so for these parameters results are only shown for the original model. The bounds of the uniform prior distributions for each parameter are the upper and lower limits on the $x$-axes.}
    \label{fig:reduced marginals}
\end{figure}

\begin{figure}[H]
    \centering
    \includegraphics[width=\textwidth]{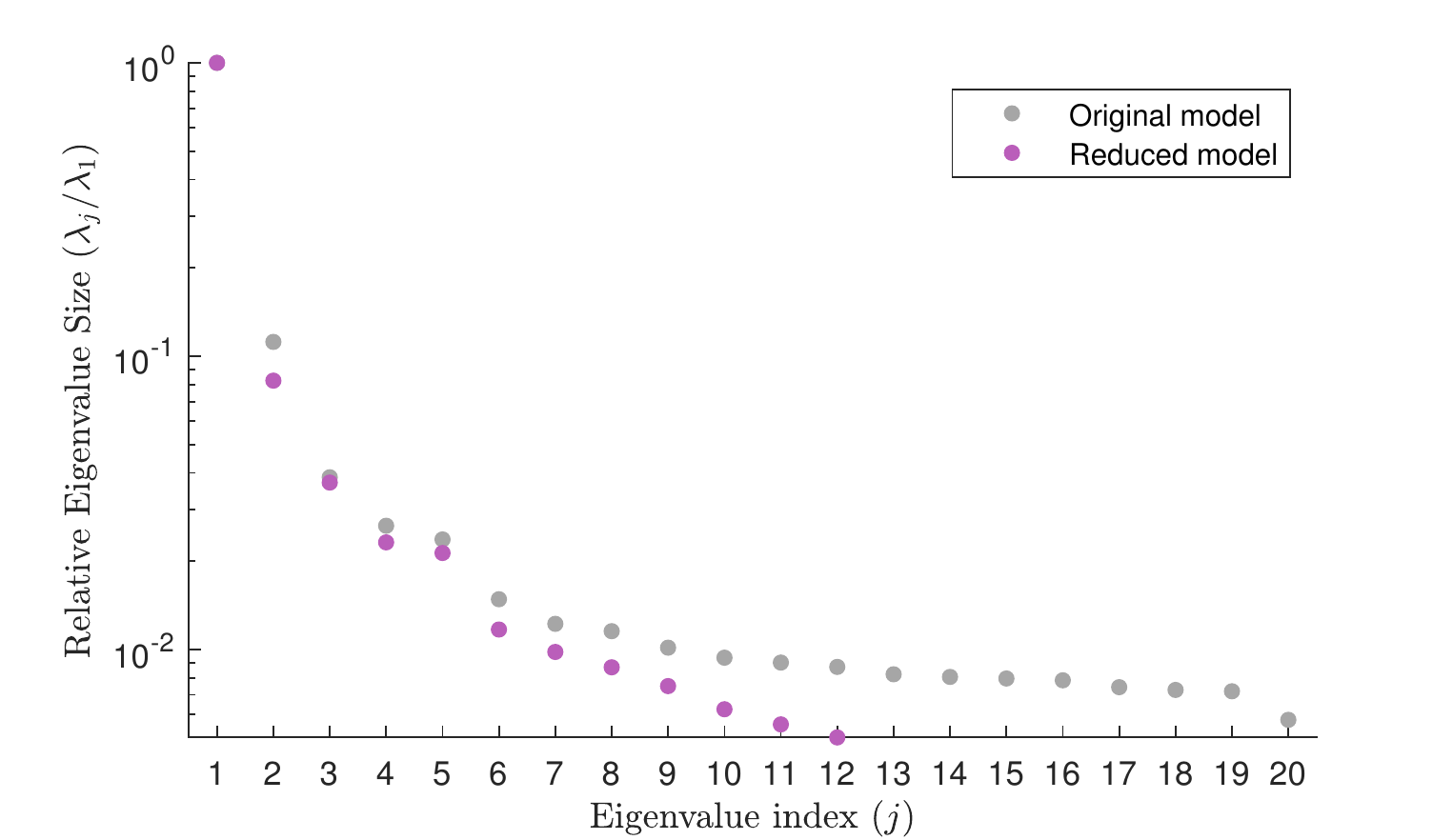}
    \caption{Relative eigenvalue sizes when compared to the leading eigenvalue. This posterior covariance analysis of sloppiness was conducted on 5000 posterior samples for each of the models. For both the original and reduced model, the relative eigenvalue sizes of independent samples produced similar results so only one sample is shown for each model here. }
    \label{fig:relative importance PCA reduced}
\end{figure}

\begin{figure}[H]
    \centering
    \includegraphics[width=\textwidth,trim={0cm 9.7cm 6.2cm 0cm},clip]{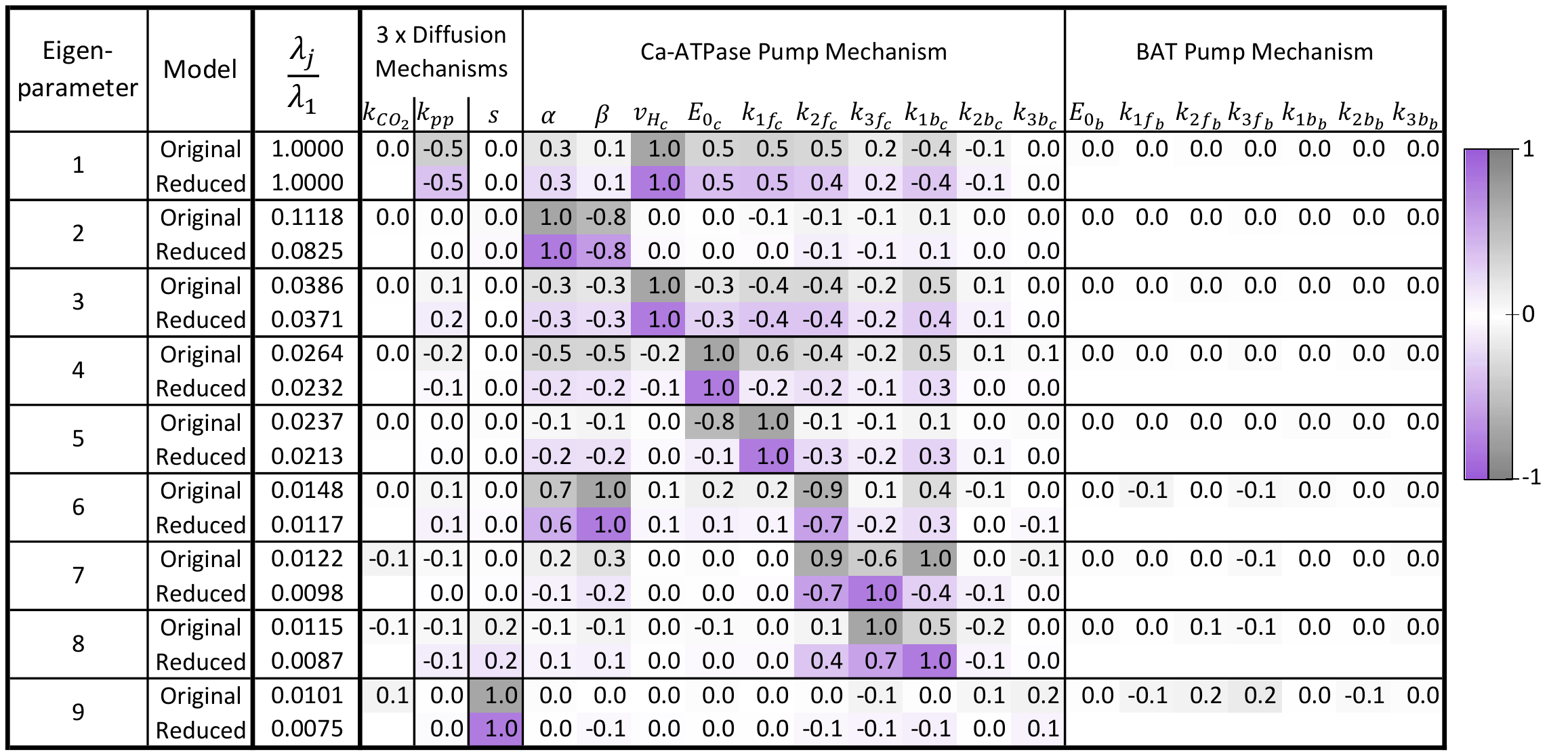}
    \caption{Eigenvector values for eigenparameters identified using a posterior covariance analysis of sloppiness, for both the original model (in grey) and reduced model (in purple). These eigenparameters correspond to the nine largest eigenvalues, and so are ordered from highest relative importance to lowest. Results beyond the ninth eigenparameter were inconsistent for the original model (Figure \ref{fig: PCA reproducibility}), and so are not shown in this comparison. See the caption of Figure \ref{fig: PCA reproducibility} for further interpretation of this figure.
}
    \label{fig:PCA reduced}
\end{figure}

In this section, we compare the original coral calcification model proposed by \cite{Galli_2018} to the reduced version which excludes both the BAT pump mechanism and the coelenteron-ECM transcellular diffusion (characterised by parameter $k_{CO2}$). Firstly, Figure \ref{fig:reduced marginals} demonstrates that the estimated marginal densities of each of the remaining parameters appear to be similar to that of the original model calibration. Secondly, Figure \ref{fig:relative importance PCA reduced} shows the decay in relative eigenvalue size using a posterior covariance analysis of model sloppiness. Here, the reduced model has a faster decay in relative eigenvalue importance. Finally, Figure \ref{fig:PCA reduced} shows the corresponding eigenparameters for both models. Each of the first six eigenparameters indicate that there is a similar model-data fit for both models. 

\newpage
\section*{Supplementary Material References}
\renewcommand{\bibsection}{}
\bibliographystyle{chicagoa}
\bibliography{Refs_supps}